\documentclass{aa}  

\usepackage{graphicx}
\usepackage[colorlinks=true,allcolors=blue]{hyperref} 
\usepackage{txfonts}
\usepackage{natbib}
\bibpunct{(}{)}{;}{a}{}{,}
\begin{document}

   \title{Nonparametric galaxy morphology from stellar and nebular emission with the CALIFA sample}

   %\subtitle{}

   \author{Angelos Nersesian
          \inst{1}
          \and
          Stefano Zibetti\inst{2}
          \and
          Francesco D'Eugenio\inst{3,4}
          \and 
          Maarten Baes\inst{1}
          }
    %\fnmsep\thanks{Just to show the usage of the elements in the author field}

   \institute{Sterrenkundig Observatorium Universiteit Gent, Krijgslaan 281 S9, B-9000 Gent, Belgium\\
              \email{angelos.nersesian@ugent.be} \and
              Osservatorio Astrofisico di Arcetri Largo Enrico Fermi 5, I-50125 Firenze, Italy \and
              Kavli Institute for Cosmology, University of Cambridge, Madingley Road, Cambridge, CB3 0HA, United Kingdom \and
              Cavendish Laboratory - Astrophysics Group, University of Cambridge, 19 JJ Thomson Avenue, Cambridge, CB3 0HE, United Kingdom
             }

   \date{Received 23 January 2023; accepted 3 March 2023}

% \abstract{}{}{}{}{} 
% 5 {} token are mandatory
 
  \abstract
   {}
   {We present a nonparametric morphology analysis of the stellar continuum and nebular emission lines for a sample of local galaxies. We explore the dependence of the various morphological parameters on wavelength and morphological type. Our goal is to quantify the difference in morphology between the stellar and nebular components.}
   {We derive the nonparametric morphological indicators of 364 galaxies from the CALIFA Survey. To calculate those indicators, we apply the \texttt{StatMorph} package on the high-quality integral field spectroscopic data cubes, as well as to the most prominent nebular emission-line maps, namely [\ion{O}{iii}]$\lambda$5007, H$\alpha$, and [\ion{N}{ii}]$\lambda$6583.}
   {We show that the physical size of galaxies, M$_{20}$ index, and concentration have a strong gradient from blue to red optical wavelengths. We find that the light distribution of the nebular emission is less concentrated than the stellar continuum. A comparison between the nonparametric indicators and the galaxy physical properties revealed a very strong correlation of the concentration with the specific star-formation rate and morphological type. Furthermore, we explore how the galaxy inclination affects our results. We find that edge-on galaxies show a more rapid change in physical size and concentration with increasing wavelength due to the increase in optical free path.}
   {We conclude that the apparent morphology of galaxies originates from the pure stellar distribution, but the morphology of the ISM presents differences with respect to the morphology of the stellar component. Our analysis also highlights the importance of dust attenuation and galaxy inclination in the measurement of nonparametric morphological indicators, especially in the the wavelength range 4000--5000~$\AA$.}


   \maketitle
%
%-------------------------------------------------------------------

\section{Introduction} \label{sec:intro}

One of the fundamental characteristics that we use to describe galaxies is their morphology, which can provide important clues about its formation and secular evolution. For example, processes like in situ star formation or merging events can be traced by quantifying the structure and morphology of a galaxy \citep{Conselice_2014ARA&A..52..291C}. The simplest method of classifying galaxies is through a visual inspection of their apparent morphology \citep{de_Vaucouleurs_1959HDP....53..275D, Sandage_2005ARA&A..43..581S, Lintott_2011MNRAS.410..166L}. In that case galaxies can be segregated into four broad populations: ellipticals, lenticulars, spirals and irregulars. Of course each galaxy group can be further divided into sub-groups, e.g. spirals can be with or without a bar structure. Usually a numerical value is assigned to each class of galaxy (Hubble stage $T$) ranging from $-5$ to $+10$, with negative numbers corresponding to ellipticals and lenticulars (early-types) and positive numbers to spirals and irregulars (late-types).  

A more quantitative technique to measure galaxy structure is by fitting the surface brightness distribution of galaxies with a general \citet{Sersic_1963BAAA....6...41S} profile \citep{Peng_2010ApJ...721..193P, Simard_2011ApJS..196...11S}, described by three free parameters: the S{\'e}rsic index $n$, the effective radius enclosing half of the light within a galaxy ($R_\mathrm{e}$), and the effective surface brightness ($\mu_\mathrm{e}$). From the S{\'e}rsic index it is possible to characterise if a galaxy is disk-like ($n<2$) or bulge-dominated ($n\geq2$). Many studies have investigated the dependence of galaxy structure on wavelength by fitting S{\'e}rsic profiles in various optical and near-infrared (NIR) broadband images \citep[e.g.][]{La_Barbera_2010MNRAS.408.1313L, Kelvin_2012MNRAS.421.1007K, Haussler_2013MNRAS.430..330H, Vulcani_2014MNRAS.441.1340V}. Despite the discrete nature of broadband photometry, those studies found a smooth transition of the structural parameters with wavelength for the early-type galaxies, and a more intense change for the late-types. The reported trends can be attributed to the stellar population age and metallicity gradients, as well as dust attenuation effects. Lastly, this method was applied in the far-infrared \textit{Herschel} bands by \citet{Mosenkov_2019A&A...622A.132M}, who found a modest dependence of the S{\'e}rsic index on wavelength.

Another method to quantify galaxy morphology is by making use of nonparametric indicators. There are two commonly used sets of nonparametric properties: the Gini--M$_{20}$ indices \citep{Abraham_2003ApJ...588..218A, Lotz_2004AJ....128..163L}, and the concentration--asymmetry--smoothness ($CAS$) system \citep{Conselice_2003ApJS..147....1C}. The advantages of measuring those indicators is (1) that they do not impose any functional form on a galaxy's surface brightness distribution, and (2) that they can be applied to any galaxy image, regardless of wavelength or redshift. Therefore, measuring these parameters is ideal to capture the underlying morphology of the different galaxy components.   

Previous studies have mainly focused on the stellar distribution and the application of nonparametric indicators on optical and NIR galaxy images \citep{Lotz_2006ApJ...636..592L, Huertas_Company_2009A&A...497..743H, Holwerda_2014ApJ...781...12H, Conselice_2009MNRAS.394.1956C, Conselice_2014ARA&A..52..291C, Psychogyios_2016A&A...591A...1P, Rodriguez_Gomez_2019MNRAS.483.4140R}. Other applications include the measurement of galaxy structure from HI maps \citep{Holwerda_2011MNRAS.416.2401H, Gebek_2023MNRAS} and CO data \citep{Davis_2022MNRAS.512.1522D}. \citet{Munoz_Mateos_2009ApJ...703.1569M} extended the study of nonparametric morphology to the dust component in the far-infrared (FIR) regime, using the \textit{Spitzer}~MIPS~70~and~160~$\mu$m images. \citet{Baes_2020A&A...641A.119B} presented, for the first time, a multi-wavelength study of the nonparametric morphology of nine nearby spiral galaxies, in order to consistently measure the variation of 
galaxy structure in the UV-submm wavelength range.

The application of nonparametric morphological parameters to broadband images only allows to recover the average properties of galaxies in a specific waveband, missing significant fraction of the nuanced information of the different stellar populations and interstellar medium (ISM). Indeed, the ISM is a very complex environment where the fundamental physical processes transpire within galaxies as a consequence of cosmologic evolution. An important asset towards that goal is deciphering the spectrum of a galaxy. A single spectrum contains information not only of the age of stellar populations, but also of the star-formation rate (SFR), stellar and gas-phase metallicity, dust content, and chemical abundances. Furthermore, a spectrum can reveal physical processes that shape the ISM such as shocks from massive stellar winds, feedback from an active galactic nucleus (AGN), the strength of the stellar radiation field, and gas collisions due to merging events \citep[see review by][]{Kewley_2019ARA&A..57..511K}.

In recent years, the advancement of Integral Field Spectroscopy (IFS) has enabled the formation of large surveys of galaxies, in particular ATLAS$^\mathrm{3D}$ \citep{Cappellari_2011MNRAS.413..813C}, Calar Alto Legacy Integral Field Area \citep[CALIFA;][]{Sanchez_2012A&A...538A...8S, Sanchez_2016A&A...594A..36S, Walcher_2014A&A...569A...1W}, Mapping Nearby Galaxies at Apache Point Observatory \citep[MaNGA;][]{Bundy_2015ApJ...798....7B} and Sydney-AAO Multi-object Integral field \citep[SAMI;][]{Bryant_2015MNRAS.447.2857B} Galaxy survey. Those surveys not only increased the sample sizes and wavelength coverage of the Integral Field Units (IFUs), but also improved in spatial resolution (although the field of view still remains quite small). An interesting aspect of the available, spatially resolved spectra, that has remained fairly unexplored, is the variability of galaxy morphology as a function of wavelength. 

As previously stated, most morphology indicator studies have been based on broadband images containing a mix of stellar continuum emission and gas line emission. By using IFU data we can mask the disturbing contribution of nebular line emission, and calculate the morphological indicators from the pure stellar emission only. In particular the wavelength dependence of these pure stellar emission indicators is an interesting thing to study, as different aspects contribute to it: stellar population gradients and gradients in dust attenuation. Conversely, we can also isolate the nebular line emission and study (for the first time systematically) the morphology of the ionised gas and compare that to the morphology of the stellar emission.

The goal of this paper is to extract the nonparametric morphological statistics from IFU frames, corresponding to the stellar continuum and to different spectral features in the optical regime, for a large number of galaxies. The CALIFA \citep{Sanchez_2012A&A...538A...8S, Sanchez_2016A&A...594A..36S, Walcher_2014A&A...569A...1W} survey\footnote{\url{http://califa.caha.es/}} perfectly suits that goal. The CALIFA survey was designed to obtain spatially resolved spectra for about 600 nearby galaxies. The advantages of the CALIFA IFU data are many, but here we lay the most important ones for our study: (1) it has a large field of view which covers the outskirts of galaxies; and (2) it includes galaxies of all Hubble types (ellipticals, lenticulars, spirals, and irregulars) and stellar masses ($\sim 10^9$ to $10^{12}~\mathrm{M}_\odot$) \citep{Walcher_2014A&A...569A...1W}. 

A layout of the paper is given here: Sec.~\ref{sec:dataset} presents the data and the galaxy sample. In Sec.~\ref{sec:methods} we describe our methodology to calculate the nonparametric morphology indicators from the IFS data cubes, and in Sec.~\ref{sec:results} we present the results of our analysis. Finally, in Sec.~\ref{sec:discussion} we try to interpret the results and in Sec.~\ref{sec:conclusions} we summarise the conclusions of this study. 

\begin{figure*}[t]
    \centering
    \includegraphics[width=16cm]{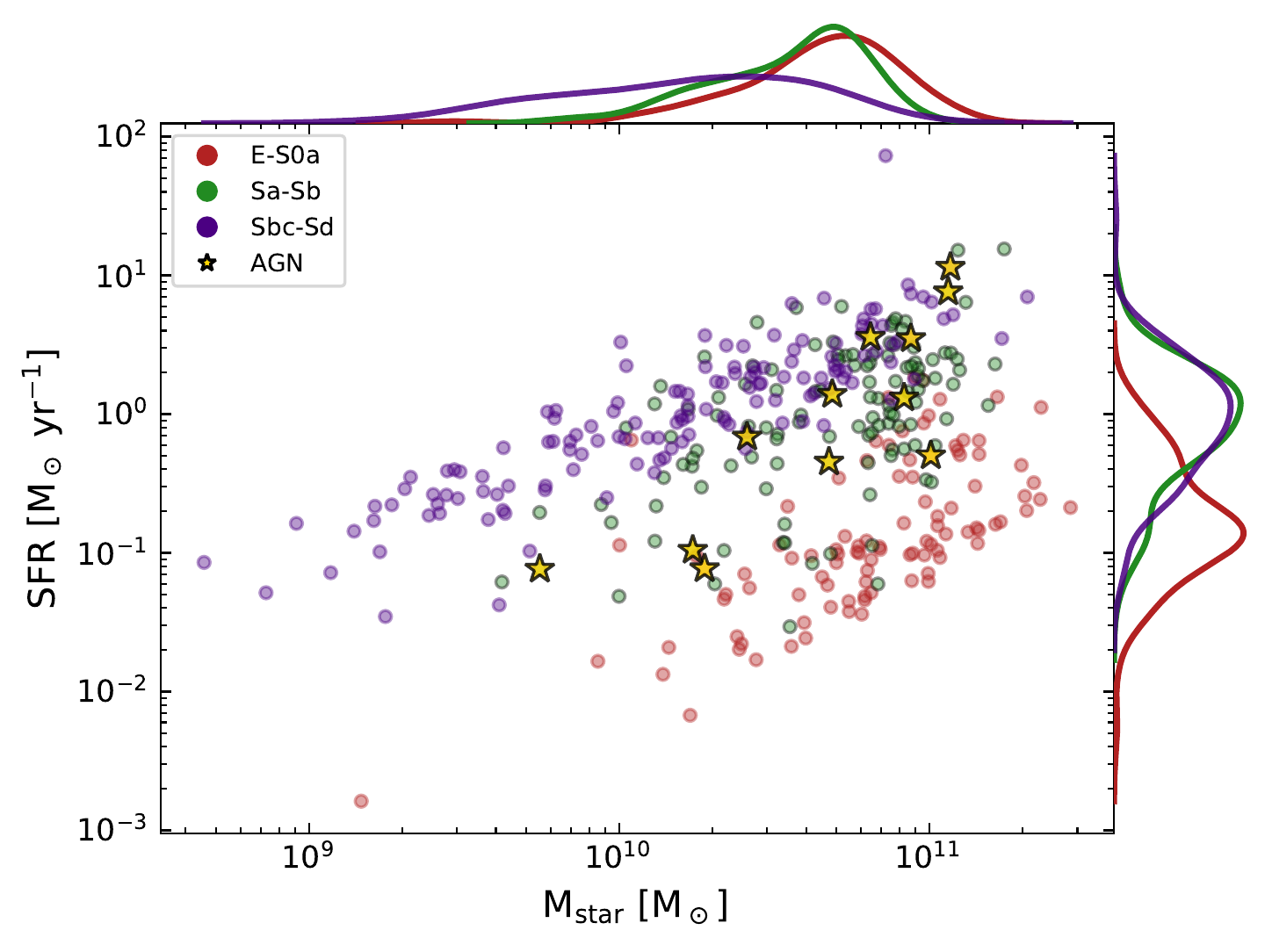}
    \caption{Scatter plot of the relation between SFR versus stellar mass. Galaxies are colour-coded according to their morphological type, and divided into three broad morphological groups as indicated in the legend of the figure. AGN galaxies are marked with golden stars. The normalised distributions of each galaxy population are shown on the top and right sides of the figure.}
    \label{fig:califa_main_sequence}
\end{figure*}

\section{Data and sample} \label{sec:dataset}

We obtained the publicly available data cubes from the third and final data release (DR3) of the CALIFA survey \citep{Sanchez_2012A&A...538A...8S, Sanchez_2016A&A...594A..36S, Walcher_2014A&A...569A...1W}. The CALIFA galaxies were observed with two different spectral gratings, the V500 (3745--7500~$\AA$) with a full-width at half maximum (FWHM) spectral resolution of $\sim 6$~$\AA$, and the V1200 (3400--4840~$\AA$) with 2.3~$\AA$. Due to the higher spectral resolution of the V1200 grating more observational time was required, therefore not all galaxies were observed in that setup. For those galaxies that were observed in both spectral gratings, a combined data cube was produced called COMBO in order to reduce any vignetting effects on the original data. The COMBO cubes by design have the same spectral resolution as the V500 and cover the 3700--7300~$\AA$ wavelength range, with a spatial sampling of 1~arcsec/spaxel. Each COMBO data cube contains 1900 wavelength elements, and information on the 1--$\sigma$ uncertainty of each pixel. For the observing strategy, data reduction and calibration, we refer to \citet{Sanchez_2012A&A...538A...8S, Sanchez_2016A&A...594A..36S}, \citet{Husemann_2013A&A...549A..87H}, and \citet{Garcia_Benito_2015A&A...576A.135G}.

Concerning our working sample, we selected only those galaxies with the COMBO data cube, so that all galaxies have a common wavelength coverage. The final sample contains 364 galaxies, with a redshift range of $0.005 < z < 0.03$, and stellar mass range of $10^{8.7}-10^{11.5}$~M$_\odot$. Out of those galaxies: 52 are of E type, 47 are of S0 and S0a types, 62 are Sa and Sab, 62 are Sb, 62 are Sbc, 60 are Sc and Scd, and 19 are of Sd type. Throughout the manuscript we categorised the galaxies into three main morphological groups: E--S0a (99 galaxies), Sa--Sb (124), and Sbc--Sd (141). 35\% of the galaxies in our sample have an inclination larger than $80\degr$, while 6\% of the galaxies are experiencing a merging event. Finally, \citet{Lacerda_2020MNRAS.492.3073L} identified the galaxies in the CALIFA survey that may host an active galactic nucleus (AGN). By cross-matching their sample with ours, we find that 12 galaxies harbour an AGN (3.3\% of the sample). Out of these 12 galaxies, 11 are classified as Type-II AGN, and only one is a Type-I AGN candidate. We decided not to exclude the AGN galaxies, since we want to explore if their morphological characteristics differ from those of normal galaxies.  

The SFRs and stellar masses were retrieved from \citet{Sanchez_2017MNRAS.469.2121S}. \citet{Sanchez_2017MNRAS.469.2121S} estimated the integrated SFRs of the CALIFA sample by applying the \citet{Kennicutt_1998ARA&A..36..189K} calibration on the H$\alpha$ luminosity maps. Furthermore, \citet{Sanchez_2017MNRAS.469.2121S} estimated the stellar masses by spatially binning the CALIFA data cubes to reach a S/N of 50, and by fitting a stellar population model to each one of those new spaxels. Then, the best-fit model within each spatial bin was used to derive the stellar-mass densities, and the integrated stellar masses of the CALIFA galaxies. A Salpeter initial mass function (IMF) \citep{Salpeter_1955ApJ...121..161S} was adopted for the derivation of both the SFR and $M_\mathrm{star}$.

Figure~\ref{fig:califa_main_sequence} displays the relationship between the SFR and $M_\mathrm{star}$. Galaxies are colour coded according to their morphological classification. A tight relation between the $M_\mathrm{star}$ and the current SFR exists for the majority of late-type spirals (blue points), establishing the star-forming sequence of galaxies \citep[e.g.][]{Brinchmann_2004MNRAS.351.1151B, Noeske_2007ApJ...660L..43N, Elbaz_2007A&A...468...33E, Whitaker_2012ApJ...754L..29W, Speagle_2014ApJS..214...15S, Tomczak_2016ApJ...817..118T}. Elliptical and lenticular galaxies (red points) are clustered just below the star-forming sequence, indicative of the cessation of star-formation activity in those systems. The early-type spirals (green points) fall somewhere in between those two sequences, having a larger scatter. AGN-host galaxies (golden stars) have lower SFRs compared to non-AGN star-forming galaxies of the same stellar mass, in agreement with previous studies \citep[e.g.][]{Silverman_2008ApJ...675.1025S, Mullaney_2015MNRAS.453L..83M, Sanchez_2018RMxAA..54..217S, Lacerda_2020MNRAS.492.3073L}. From this plot we can see that the galaxies in the CALIFA sample cover a representative volume of this parameter space relative to galaxies in the Local Universe.

\begin{figure*}[ht!]
    \centering
    \includegraphics[width=\textwidth]{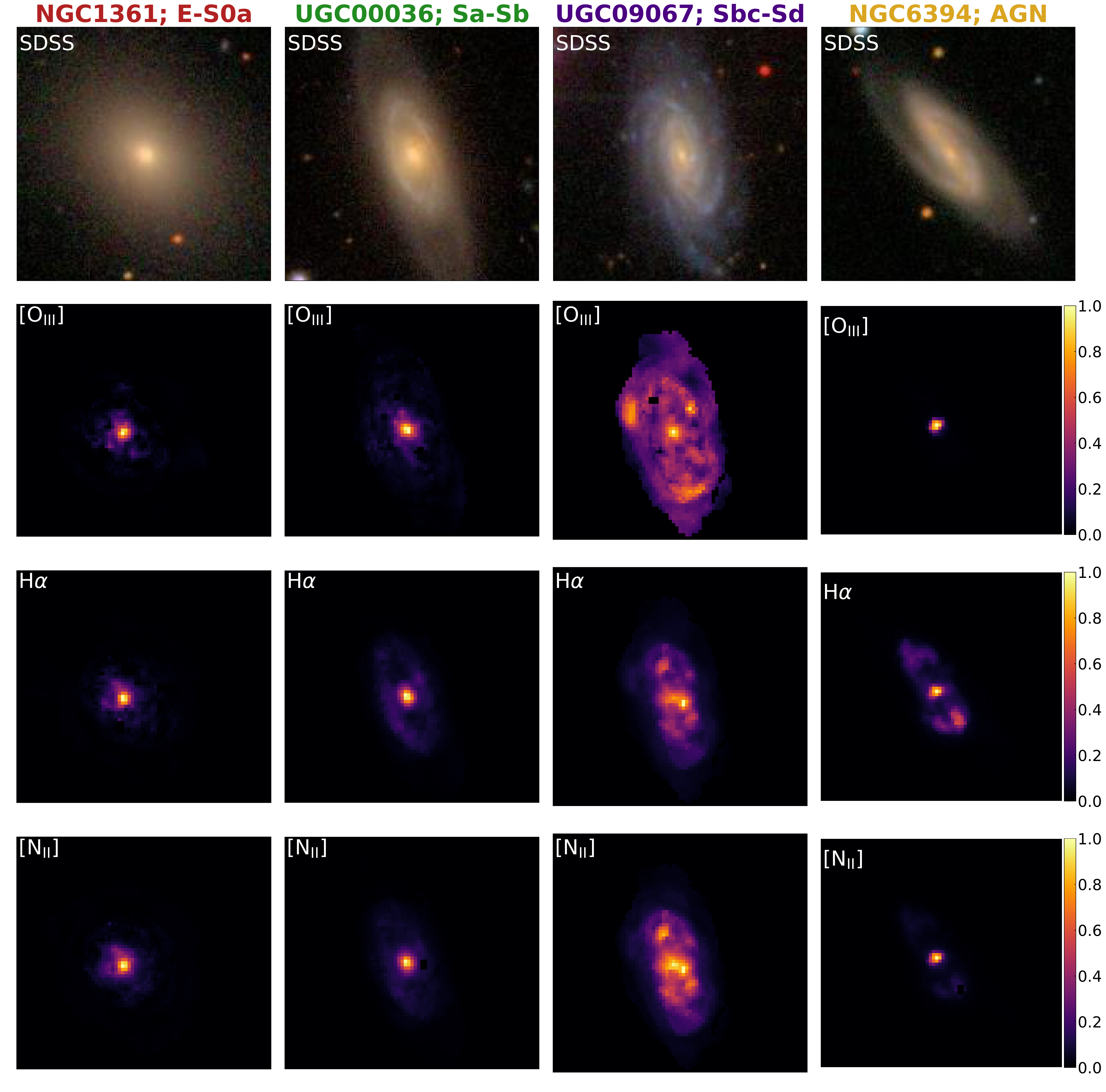}
    \caption{Individual emission-line maps of four galaxies in our sample. Each galaxy was randomly selected from its corresponding morphological group. We also show an AGN host galaxy (fourth column). For each galaxy, we show an optical image from SDSS (top row), and the continuum-subtracted emission-line maps of [\ion{O}{iii}]$\lambda$5007, H$\alpha$, and [\ion{N}{ii}]$\lambda$6583. The emission-line maps are in units of normalised flux density.}
    \label{fig:thumbnail}
\end{figure*}

\section{Method of analysis} \label{sec:methods}

There are six main morphological properties that we consider in this study, namely the half-light radius ($R_\mathrm{half}$), the Gini and M$_{20}$ statistics, the concentration ($C$), the asymmetry ($A$), and the smoothness ($S$) parameters. We give a brief definition for each one of these parameters.

The half-light radius is often used to characterise the physical size of a given galaxy. It is defined as the elliptical aperture of the isophote that contains half of the emitted light in a particular wavelength. It has been shown that the size of a galaxy varies with wavelength \citep[e.g.][]{Kelvin_2012MNRAS.421.1007K, Vulcani_2014MNRAS.441.1340V, Baes_2020A&A...641A.119B}. The change in size may be related to the presence of different stellar or dust components, hence revealing information of the internal structure of galaxies. 

\citet{Lotz_2004AJ....128..163L, Lotz_2008ApJ...672..177L} measured the Gini--M$_{20}$ statistics and introduced a new method to segregate galaxies according to their morphological type and merging status. The Gini coefficient is a quantity traditionally used in economics to measure wealth inequality. However, \citet{Abraham_2003ApJ...588..218A} introduced the Gini coefficient and applied it to galaxy images. The Gini index indicates the spread in the pixel values within a predefined aperture. In other words, the Gini index provides an estimate of the concentration of light in the brightest pixels in galaxies with an arbitrary shape. The Gini index is defined as:

\begin{equation} \label{eq:gini}
\mathrm{Gini} = \frac{1}{\overline{f}~k~(k-1)} \sum_{i=1}^{k} (2i - k - 1)~f_i,
\end{equation}

\noindent where $\overline{f}$ is the mean flux over the pixel values, $f_i$ is the flux of the $i^\mathrm{th}$ pixel, and $k$ is the total number of pixels assigned to a galaxy. First, the $k$ pixel values are sorted from minimum to maximum. Then, the pixels are divided equally into 50\% bright and 50\% faint values. When the Gini index is calculated for the bright (faint) pixels it takes positive (negative) values. The final value of the Gini index is calculated by summing up the difference between the brightest and faintest pixels, the second brightest and second faintest pixels, etc, and by dividing the sum by $\overline{f}~k~(k-1)$. The final range of the Gini index is between 0 and 1. When Gini~$=0$ then a galaxy has a uniform flux distribution. Conversely, Gini~$=1$ if the light of the galaxy is concentrated in just few pixels.

Similarly, M$_{20}$ is another measure of concentration in galaxies, but it is more sensitive to the brightest regions outside the centre of the galaxy. In particular, M$_{20}$ represents the logarithm of the second moment of the brightest 20\% of a galaxy relative to the total second-order central moment, M$_\mathrm{tot}$. The M$_\mathrm{tot}$ is calculated through:

\begin{equation} \label{eq:m20_1}
\mathrm{M}_\mathrm{tot} = \sum_{i=1}^{k} \mathrm{M}_i = \sum_{i=1}^{k} f_i \left[\left(x_i-x_c\right)^2+\left(y_i-y_c\right)^2\right],
\end{equation}

\noindent where $f_i$ is the flux of pixel ($x_i$, $y_i$) and ($x_c$, $y_c$) are the coordinates of the galaxy’s centre. M$_{20}$ is given then by:

\begin{equation} \label{eq:m20_2}
\mathrm{M}_\mathrm{20} = \log\left(\frac{\sum_{i}~\mathrm{M}_i}{\mathrm{M}_\mathrm{tot}}\right),~\mathrm{while}~\sum_{i=1}^{k} f_i < 0.2~f_\mathrm{tot},
\end{equation}

\noindent where $f_\mathrm{tot}$ is the total flux of the pixels.

In general, M$_{20}$ is a suitable parameter to infer the structural properties of galaxies such as spiral arms, bars, rings, tidal perturbations, and multiple nuclei \citep{Lotz_2004AJ....128..163L, Holwerda_2014ApJ...781...12H}.

The $CAS$ system can be used to quantify consistently and objectively the morphological parameters of galaxies. \citet{Conselice_2003ApJS..147....1C} argued that the $CAS$ system has the potential of being used as an alternate classification scheme for galaxies, one that is more tightly connected to the physical properties. As its name suggests, the concentration index measures the concentration of the light distribution in galaxies. It is defined as:

\begin{equation} \label{eq:con}
C = 5~\log\left(\frac{R_{80}}{R_{20}}\right),
\end{equation}

\noindent with $R_{20}$ and $R_{80}$ being the elliptical radii of the isophotes that contain 20\% and 80\% of the light, respectively. A reasonable correlation has already been found for the concentration of the light distribution with Hubble stage \citep{Morgan_1958PASP...70..364M, Morgan_1959PASP...71..394M, Okamura_1984ApJ...280....7O, Bershady_2000AJ....119.2645B}, bulge-to-disk ratio \citep{Conselice_2003ApJS..147....1C}, and supermassive black hole mass \citep{Aswathy_2018MNRAS.477.2399A}.

The asymmetry index is simply measured by subtracting the galaxy image frame, rotated by $180\degr$, from the original image.

\begin{equation} \label{eq:asym}
A = \frac{\sum_{i,j}~\lvert f_{ij} - f_{ij}^{180\degr} \rvert}{\sum_{i,j}~\lvert f_{ij} \rvert} - A_\mathrm{bgr},
\end{equation}

\noindent where $A_\mathrm{bgr}$ is the mean asymmetry value of the background, $f_{ij}$ are the pixel fluxes of the original image, and $f_{ij}^{180\degr}$ are the pixel fluxes of the rotated image. 

It has been shown that the asymmetry index of star-forming galaxies at optical wavelengths correlates reasonably well with optical broadband colour \citep{Conselice_2000ApJ...529..886C, Conselice_2003ApJS..147....1C}. However, the primary use of this indicator is tracing galaxy mergers and interactions, which often cause strong asymmetries.

Lastly, the smoothness index is obtained by subtracting a smoothed version of the galaxy image frame from the original image. A boxcar of a certain width (0.25 the Petrosian radius) is used to smooth the galaxy’s original image \citep{Lotz_2004AJ....128..163L}. Smoothness is defined as:

\begin{equation} \label{eq:smooth}
S = \frac{\sum_{i,j}~\lvert f_{ij} - f_{ij}^\mathrm{S} \rvert}{\sum_{i,j}~\lvert f_{ij} \rvert} - S_\mathrm{bgr},
\end{equation}

\noindent where $\mathrm{S}_\mathrm{bgr}$ is the mean smoothness of the background, $f_{ij}$ are the pixel fluxes of the original image, and $f_{ij}^\mathrm{S}$ are the pixel fluxes of the smoothed image.

Galaxies that have large smoothness values have a more clumpy appearance. That is why the smoothness index is also referred to as the clumpiness index. In principal, this indicator is ideal to segregate between quiescent early-type galaxies and actively star-forming late-type galaxies. Yet, the smoothness index is weakly correlated with SFR \citep{Conselice_2003ApJS..147....1C}.

\begin{figure*}
    \centering
    \includegraphics[width=\textwidth]{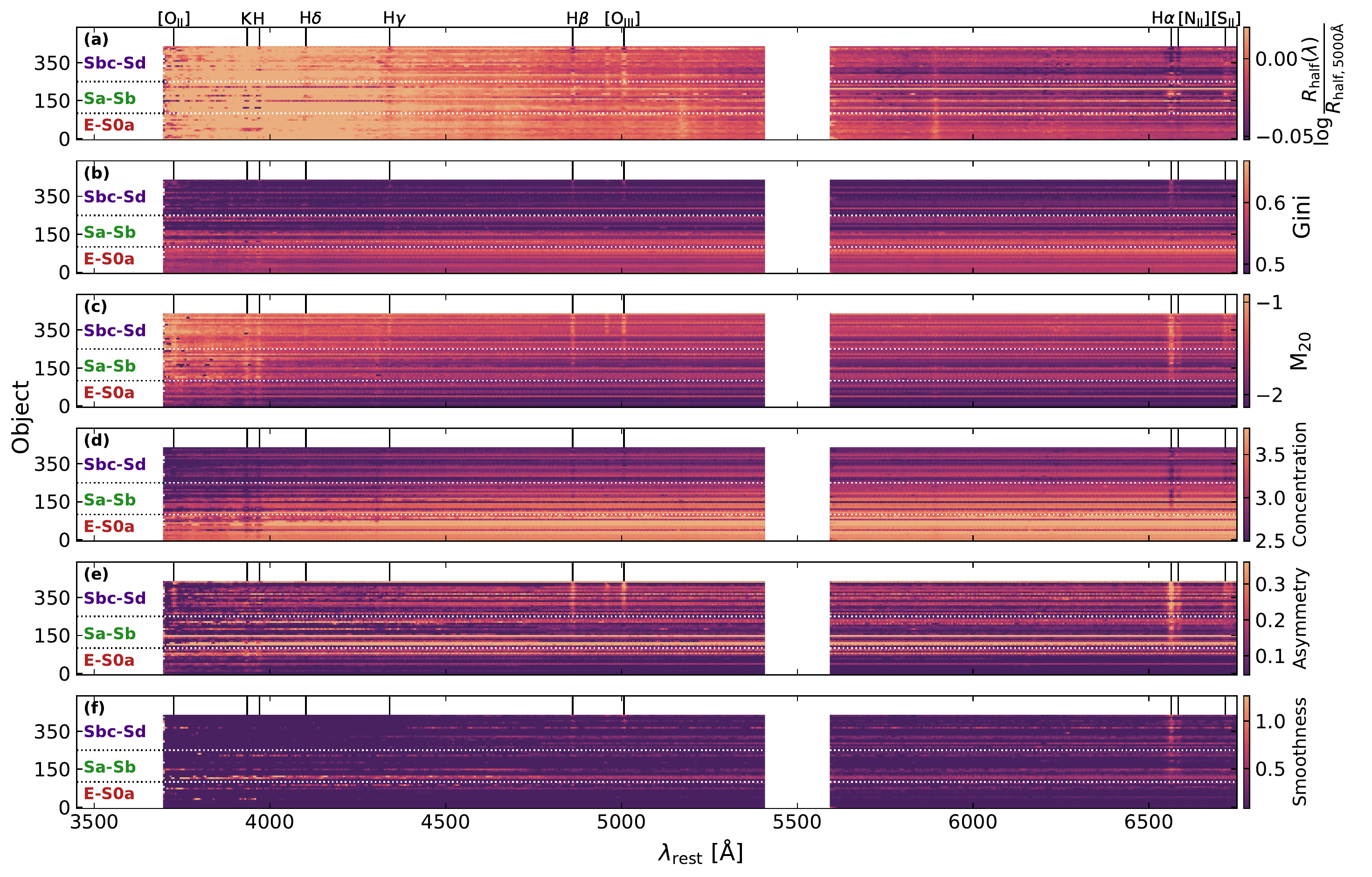}
    \caption{Nonparametric morphological properties as a function of rest-frame wavelength. In this plot we show the results of \texttt{Statmorph} applied on the IFU data of 364 CALIFA galaxies. From top to bottom we present the wavelength dependence of $\log R_\mathrm{half}$($\lambda$)/$R_\mathrm{half, 5000\AA}$, Gini, M$_{20}$, $C$, $A$, and $S$. The galaxy spectra are stacked according to their morphological type, and they are sorted with increasing SFR per galaxy population. The horizontal dotted lines show the borders between the three galaxy populations. The vertical solid lines indicate the central wavelength of the most prominent spectral features. The white rectangle centred around 5500~$\AA$ masks the area of the telluric features in the spectrum. The spectra were binned in wavelength to smooth out the noise.}
    \label{fig:morph_spectra}
\end{figure*}

We use the \texttt{python} package \texttt{StatMorph} \citep{Rodriguez_Gomez_2019MNRAS.483.4140R} to measure the morphological statistics of the CALIFA galaxies. \texttt{StatMorph} was build upon the previous studies by \citet{Lotz_2004AJ....128..163L, Lotz_2006ApJ...636..592L, Lotz_2008ApJ...672..177L, Lotz_2008MNRAS.391.1137L} for calculating the Gini--M$_{20}$ statistics, and the $CAS$ system indices \citep{Conselice_2000ApJ...529..886C} of a given image. Specifically, we apply \texttt{StatMorph} on each frame included in a CALIFA IFS data cube to retrieve the aforementioned morphological parameters. Of course, the IFU frames contain information on various nebular emission lines as well as absorption features by the stellar continuum. We are interested in measuring if there is a change in morphological structure for different nebular emission lines, and compare with the morphology of the stellar continuum. 

In order to retrieve the morphological parameters of emission lines it is imperative to subtract the stellar continuum and integrate the frames belonging to each emission line. \citet{Zibetti_2017MNRAS.468.1902Z} provided data cubes with the maps of 25 continuum-subtracted emission lines. To decouple the emission lines from the stellar continuum, the data cubes were preprocessed with an adaptive-kernel smoothing algorithm \citep[see][for more details]{Zibetti_2009arXiv0911.4956Z, Zibetti_2009MNRAS.400.1181Z, Zibetti_2017MNRAS.468.1902Z}. The standard \texttt{pPXF}+\texttt{GANDALF} fitting procedure \citep{Cappellari_2004PASP..116..138C, Sarzi_2006MNRAS.366.1151S} was then applied to the smoothed cubes. Note that the adaptive smoothing applied to the CALIFA cubes results in a degradation of the spatial resolution at low surface brightness. However, for the adopted detection threshold in this work, the smoothing does not result in a substantial loss of resolution relative to the native PSF of the CALIFA data cubes in most of the cases. In our analysis, we take into account the corresponding noise maps included in the CALIFA IFS and continuum-subtracted emission line data cubes. Figure~\ref{fig:thumbnail} presents an optical $gri$ broad-band image from SDSS and the individual continuum-subtracted emission line maps of [\ion{O}{iii}]$\lambda$5007, H$\alpha$, and [\ion{N}{ii}]$\lambda$6583, of four CALIFA galaxies in our sample. Each galaxy was randomly selected from their corresponding morphological group. It is already evident from this plot that the morphology of the nebular component has some differences with respect to the stellar body of the galaxies.

For each galaxy we created a segmentation map \citep{Lotz_2004AJ....128..163L} based on a frame that traces the stellar continuum emission with a detection threshold of 1--$\sigma$ noise level. The segmentation map is later used for the analysis of all wavelength frames (1900 in total), of a particular galaxy. Regarding the individual nebular emission maps, we calculated the morphological statistics for a range of detection thresholds of their corresponding segmentation map (0.50--3.50~$\sigma$ noise level). From those measurements, we estimated the median and the $16^\mathrm{th}$--$84^\mathrm{th}$ percentile range of the morphological properties. In our analysis, we use the morphological statistics based on the segmentation map with a detection threshold of 2.50. We provide the uncertainties on the nonparametric morphological properties of the nebular emission maps by quoting the calculated $16^\mathrm{th}$--$84^\mathrm{th}$ percentile range. We also introduce a minimum error of 5\% to account for both underlying systematic effects on the production of the continuum subtracted maps and the measurement of the nonparametric indices.   

After applying \texttt{StatMorph} to the CALIFA data cubes, we performed a quality check on the results. In \texttt{StatMorph} there is a flagging system. When \texttt{flag == 0} the measurements can be trusted, while \texttt{flag == 1} indicates a problematic measurement. Furthermore, if the mean signal-to-noise ratio (S/N) is lower than 2.5, then the measurements cannot be trusted \citep{Lotz_2006ApJ...636..592L}. In the case of the main CALIFA data cubes, most of the frames, if not all, pass those two criteria. There are few problematic frames at the wavelength edges of each data cube that fail both criteria, and therefore we exclude those frames from any statistical calculation. Lastly, in the case of the individual emission lines, we excluded 37 galaxies from the statistical analysis in subsection~\ref{subsec:eline_morph_params} and subsection~\ref{subsec:eline_morph_params_phys}, because \texttt{StatMorph} raised a \texttt{flag == 1}. The majority of those 37 galaxies are E--S0a types (27 galaxies), which are expected not to have strong nebular emission lines. Indeed, a closer inspection of the individual emission line maps of the \texttt{flagged} galaxies revealed that most spaxels have a very low S/N (less than 2.5).

\begin{figure}[t]
    \centering
    \includegraphics[width=\columnwidth]{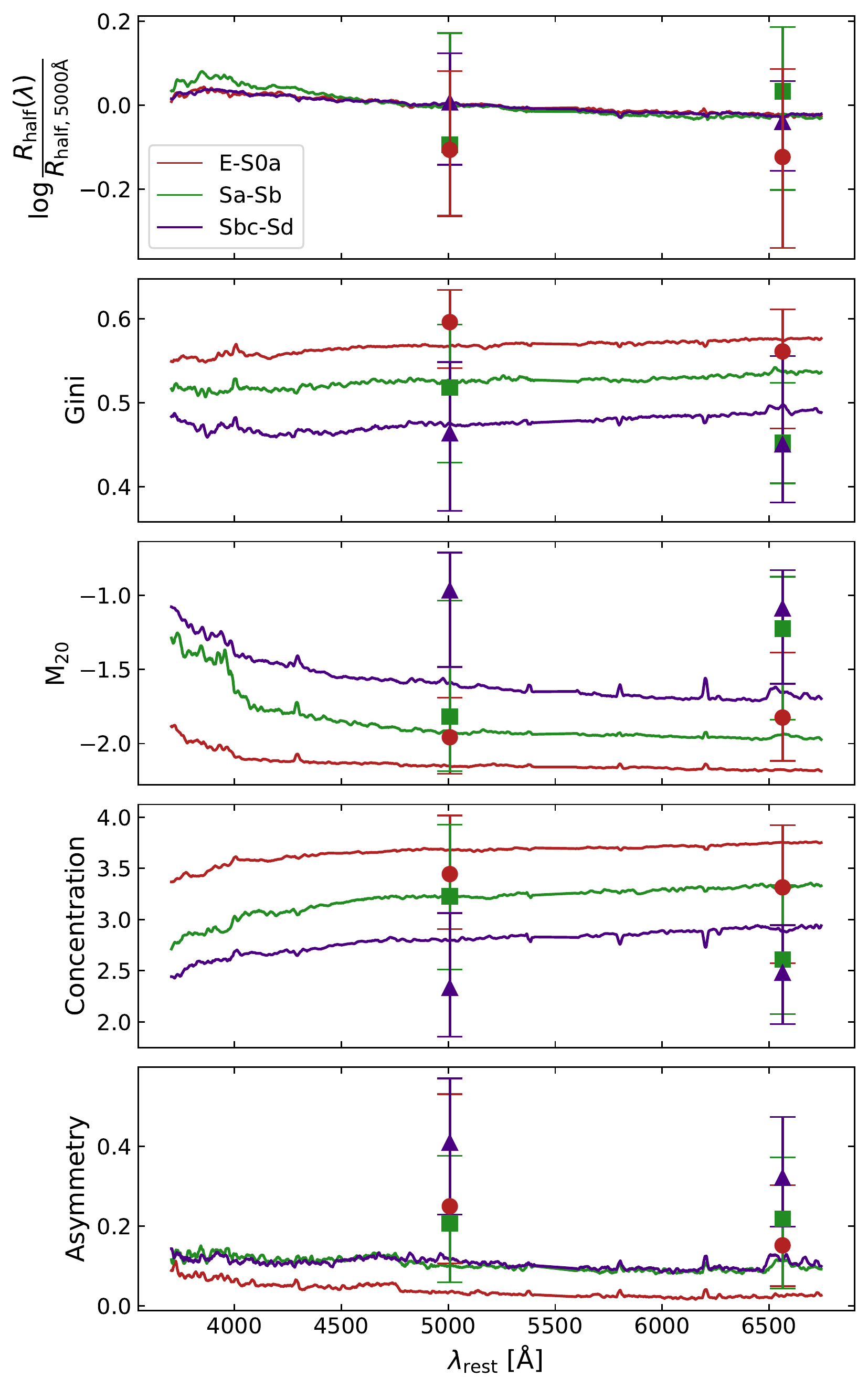}
    \caption{Median trends of the nonparametric morphological properties as a function of rest-frame wavelength. We bin the galaxies according to their morphological type. The coloured markers represent the median values of the nonparametric morphological properties for each galaxy group, as measured from the emission line maps of [\ion{O}{iii}]$\lambda$5007 and H$\alpha$. The error bars represent the 16$^\mathrm{th}$--84$^\mathrm{th}$ percentile range. The smoothness index was omitted from this plot due to the small number of wavelength-elements with reliable measurement.}
    \label{fig:medians}
\end{figure}

\section{Results} \label{sec:results}

In this section, we show the trends of galaxy morphological properties with wavelength. However, our main focus lies in the morphology of the nebular component of a galaxy, and how it differs from the morphology measured from the stellar component. We first consider trends in the morphological parameters as function of wavelength (subsection~\ref{subsec:ifu_vs_wav}). Later on, we explore the M$_{20}$--Gini relation of the stellar continuum and of various individual spectral features (subsection~\ref{subsec:eline_morph_params}). 

\begin{figure*}[t]
    \centering
    \includegraphics[width=\textwidth]{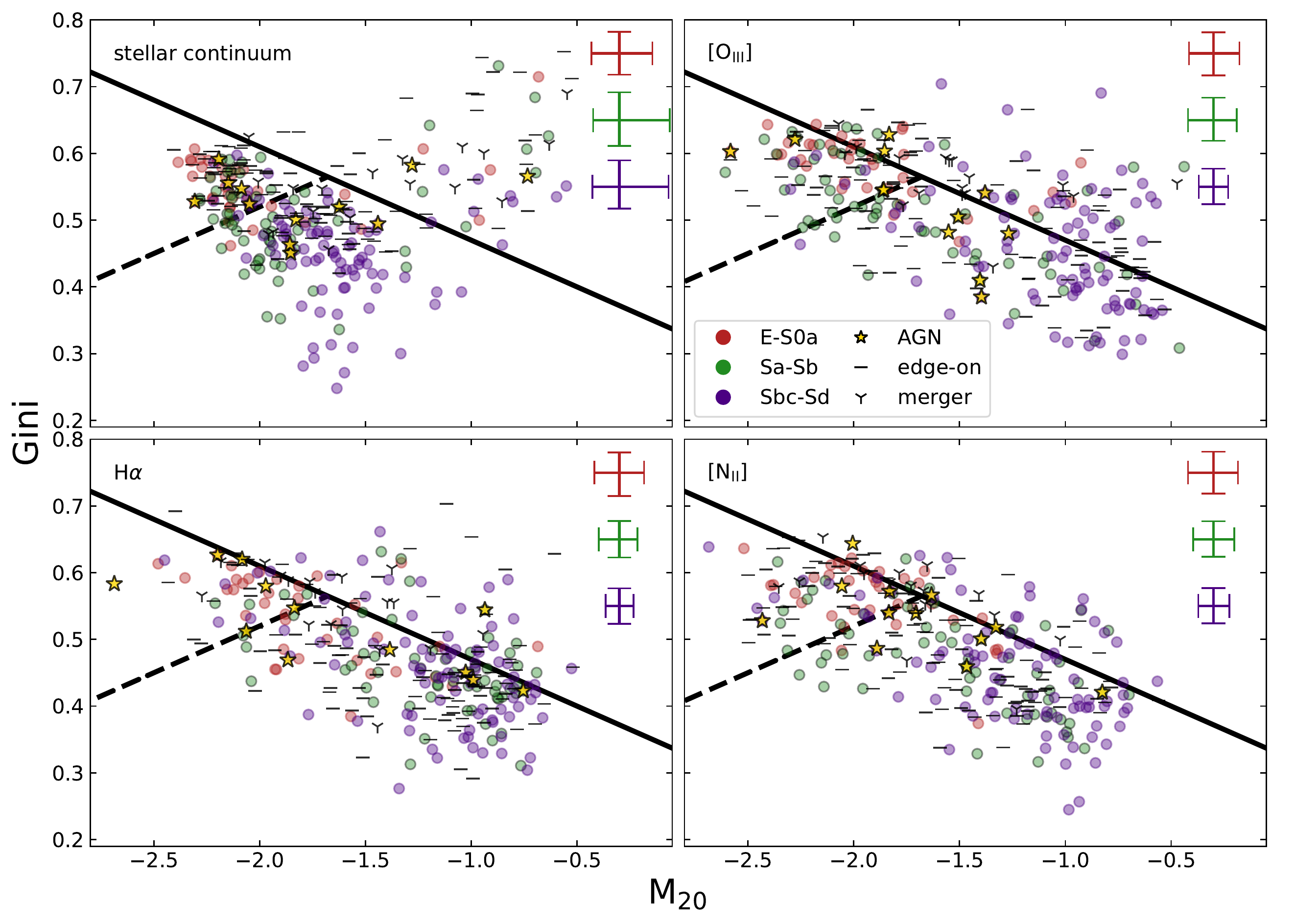}
    \caption{M$_{20}$--Gini relation for the CALIFA galaxies. Each panel shows the scatter plot of the M$_{20}$--Gini relation for the stellar continuum and three different spectral features. The solid and dashed black lines are defined in \citet{Lotz_2008ApJ...672..177L}. The solid black line roughly separates galaxies between isolated and mergers, whereas the dashed black line divides galaxies into early and late types. Each point represents a galaxy, colour-coded according to their morphological group. AGNs are marked with a golden star. All galaxies with an inclination greater than $80\degr$ are marked with a horizontal bar, while all mergers are indicated with a triangle. The average uncertainties of each galaxy group are shown in the top-right corner of each panel.}
    \label{fig:gini_m20}
\end{figure*}

\subsection{Morphological parameters vs wavelength} \label{subsec:ifu_vs_wav}

In Fig.~\ref{fig:morph_spectra} we show the spectral dependence of the main six morphological parameters retrieved by applying \texttt{Statmorph} to the CALIFA data cubes. The following quantities are displayed from top to bottom: the ratio between the half-light size measured in each wavelength over the half-light size measured at 5000~$\AA$, the Gini index, M$_{20}$, concentration ($C$), asymmetry ($A$), and smoothness ($S$). Galaxies are stacked according to their morphological type, and sorted with increasing SFR per galaxy population. The most prominent spectral features are indicated with the vertical solid lines. In Fig.~\ref{fig:medians} we show the median trends of each morphological index as a function of rest-frame wavelength. We proceed to discuss the results shown in Fig.~\ref{fig:morph_spectra}, panel-by-panel.

\paragraph{$\mathcal{R} = \log R_\mathrm{half}$($\lambda$)/$R_\mathrm{half, 5000\AA}$.} Instead of displaying the absolute size of galaxies alone, it is often more informative to present the size ratio measured at two different wavelengths. The reasoning is that a ratio like $\mathcal{R}$ can be seen as a conventional colour gradient. We chose to normalise all spectra at 5000~$\AA$ due to the flattening of the spectrum beyond that wavelength. Focusing at the top panel of Fig.~\ref{fig:morph_spectra} and the corresponding median trends in Fig.~\ref{fig:medians}, we clearly notice the dependence of galaxy size on wavelength. In all cases, a decreasing gradient can be seen from short to longer wavelengths, in agreement with the results of \citet{Kelvin_2012MNRAS.421.1007K} and \citet{Vulcani_2014MNRAS.441.1340V}. All galaxies appear to be more extended ($\mathcal{R}>0$) at blue wavelengths ($<4500~\AA$), with their size decreasing as we move towards redder wavelengths, and then remaining roughly constant ($>5000~\AA$). The change in size with wavelength seems to be more drastic for the late-type galaxies compared to the early-types. The observed gradient traces the emission of the various stellar populations. The younger, more extended, stellar disk at shorter wavelengths, and the older, more compact, stellar bulge in the longer wavelengths.

\paragraph{Gini index.} In panel (b) of Fig.~\ref{fig:morph_spectra} we show the Gini index as function of wavelength. Qualitatively the Gini index of E--S0a and for many Sa-type galaxies is larger than the Gini index of Sb and Sbc--Sd galaxies. This is the expected behaviour since higher Gini values indicate that the light distribution is concentrated in few pixels, which is true for the early-type galaxies. There is a subtle but non-negligible increasing gradient of Gini with wavelength for Sb and Sbc--Sd galaxies (see also the corresponding median trends in Fig.~\ref{fig:medians}), suggesting that late-type spirals appear less extended at longer optical wavelengths. On average, the Gini index remains roughly constant with wavelength for most galaxies (E--S0a: 0.57; Sa--Sb: 0.53; Sbc--Sd: 0.48).

\paragraph{M$_{20}$ index.} The strongest variation with wavelength is seen for the M$_{20}$ index in panel (c) of Fig.~\ref{fig:morph_spectra} and middle panel of Fig.~\ref{fig:medians}. Early- and late-type spirals take very large M$_{20}$ values at the bluest wavelengths (<4500~$\AA$) followed by a decreasing trend towards redder wavelengths. On the other hand, E--S0a have the lowest values among the different galaxy types and remain roughly constant as a function of wavelength. This means that early-type galaxies are more concentrated across the entire optical wavelength regime, while late-type galaxies are more (less) extended at shorter (longer) wavelengths. In other words, the spiral structure is more prominent at short wavelengths. In addition, H$\beta$, [\ion{O}{III}]$\lambda$5007, H$\alpha$, and [\ion{N}{II}]$\lambda$6583 feature prominently in this panel. In late-type galaxies, M$_{20}$ is always higher for the aforementioned spectral features than for the continuum, regardless of their wavelength range. This is an interesting finding since it is an indication of the differences between the morphology of the stellar continuum and the nebular component of the ISM.  

\paragraph{Concentration.} The concentration index shares a lot of similarities with both the Gini and M$_{20}$ statistics. In fact, it has been shown that $C$ anti-correlates with M$_{20}$ and correlates with Gini \citep{Abraham_2003ApJ...588..218A, Baes_2020A&A...641A.119B}. Indeed, in panel (d) of Fig.~\ref{fig:morph_spectra} we notice that for the spiral galaxies $C$ increases with wavelength. Sbc--Sd galaxies are more extended ($C < 2.7$) at the shortest wavelengths (<4500~$\AA$), and appear more compact ($C > 2.8$) with increasing wavelength. E--S0a galaxies are in general more compact with a small relative change in concentration at all optical wavelengths. Similarly to M$_{20}$, spectral features are visible but they are less pronounced. We note here that M$_{20}$ is much more sensitive to the morphology of emission lines than the other nonparametric morphological indices. 

\paragraph{Asymmetry.} In panel (e) of Fig.~\ref{fig:morph_spectra} we illustrate the trends of the asymmetry parameter with wavelength. The asymmetry parameter of each galaxy remains roughly constant across the wavelength range (as it is also seen for the corresponding median trend in Fig.~\ref{fig:medians}). Most, if not all, E--S0a galaxies are more symmetric than the late-type galaxies. This is the expected behaviour as late-type galaxies are actively star-forming, resulting in a very complex internal structure with significant asymmetric features \citep{Hambleton_2011MNRAS.418..801H}. For spiral galaxies, the emission lines in panel (e) and in particular H$\beta$, [\ion{O}{III}]$\lambda$5007, and H$\alpha$ display high asymmetry values. Notwithstanding, a lot of the asymmetry signal at the wavelengths corresponding to the most prominent emission lines is due to a kinematic effect caused by the rotation, rather than by structure. The blue/redshift that affects the approaching/receding side of the galaxy results in a very strong $180\degr$-asymmetry when looking at individual wavelength frames (i.e. not looking at the emission maps). The flux distribution is only weakly asymmetric once integrate and remove the Doppler shift. In general, the accuracy of the measurement of the $A$ is sensitive to the spatial resolution and S/N \citep{Lotz_2004AJ....128..163L, Povic_2015MNRAS.453.1644P}. The asymmetry parameter can be reliably measured when $S/N>50$. The spatial resolution is crucial too, since objects will appear more symmetric when they are poorly resolved \citep{Conselice_2000ApJ...529..886C}.   

\paragraph{Smoothness.} Smoothness is the least reliable parameter in our analysis. Similarly to asymmetry, the smoothness parameter is prone to noise effects especially for ground-based observations \citep{Lotz_2004AJ....128..163L, Povic_2015MNRAS.453.1644P}. It also requires a sufficient image resolution \citep[i.e. lower than 1000~pc][]{Lotz_2004AJ....128..163L}, in order to detect if a disk is clumpy or smooth and unfortunately the CALIFA survey does not provide the required spatial resolution (on average 1055~pc). Many of the wavelength-elements of the smoothness index return either values very close to zero or values that are entirely unreliable. In the bottom panel of Fig.~\ref{fig:morph_spectra}, we only show the smoothness values with a reliable measurement, while we chose to omit the median trends of smoothness from Fig.~\ref{fig:medians}. From the bottom panel of Fig.~\ref{fig:morph_spectra}, one can notice that all galaxies appear to be smooth across the wavelength range with Sa galaxies to be more clumpy-like. Interestingly, for the higher S/N dataframes like the H$\alpha$ emission line and the surrounding pixels, we are able to detect that the Sb and Sbc--Sd galaxies have more clumpy ISM, since H$\alpha$ emission originates primarily from star-forming regions. For the rest of the analysis we are not going to discuss $S$ due to the poor image resolution, and because we are using the continuum-subtracted emission line maps which are smoothed with a kernel function, smearing out any resolved clumpy regions.

In Fig.~\ref{fig:medians}, we also show the median values of each galaxy group and for each nonparametric morphological index, as measured from the emission line maps of [\ion{O}{iii}]$\lambda$5007 and H$\alpha$. We immediately notice the significant differences between the morphology of the ionised ISM and the stellar continuum emission. In general, the median values of the emission lines for each morphological property and galaxy group, broadly follow the median trends of the corresponding stellar continuum emission, i.e. E--S0a galaxies appear to be more concentrated and symmetric than Sbc--Sd galaxies. The median values of Gini, M$_{20}$, and concentration as measured from the nebular lines imply that the morphology of the ionised ISM is less concentrated than the stellar continuum, while from the asymmetry parameter we notice that the nebular morphology is more asymmetric compared to the one of the stellar body. A more detailed analysis of these results follows in Sec.~\ref{subsec:eline_morph_params} and Sec.~\ref{sec:discussion}.

Once more, we would like to note here that both $A$ and $S$ suffer from large uncertainties and biases for the reasons discussed above. Hence, the results presented in panels (e) and (f) of Fig.~\ref{fig:morph_spectra} and the bottom panel of Fig.~\ref{fig:medians} must be interpreted with caution.

\subsection{Nonparametric morphology of individual emission line maps - The M$_{20}$--Gini relation} \label{subsec:eline_morph_params}

In this section we present the nonparametric morphological parameters measured for three of the most prominent spectral features in Fig.~\ref{fig:morph_spectra}, namely, the [\ion{O}{iii}]$\lambda$5007, H$\alpha$, and [\ion{N}{ii}]$\lambda$6583 emission lines. We are interested comparing the parameters of the individual emission lines with the corresponding morphology, measured from the stellar continuum. 

We start by focusing on the relation between the Gini index and M$_{20}$. For each galaxy we derive the nonparametric indicators of the stellar continuum and its corresponding uncertainties, by calculating the median and the 16$^\mathrm{th}$--84$^\mathrm{th}$ percentile range from the nonparametric morphology spectra. 

We remind the reader that we excluded 37 galaxies from the statistical analysis of this subsection, because \texttt{StatMorph} raised a \texttt{flag == 1}. After we exclude the \texttt{flagged} galaxies from our sample, we end up with: 72 E--S0a, 116 Sa--Sb, and 139 Sbc--Sd. Out of these galaxies: 12 host an AGN, 122 are edge-on, and 19 are experiencing a merging event.

We plot the M$_{20}$--Gini relation in Fig.~\ref{fig:gini_m20}. The M$_{20}$--Gini relation is primarily used to separate galaxies into early or late-type galaxies, as well as to detect mergers. The solid and dashed black lines in each panel of Fig.~\ref{fig:gini_m20} were defined by \citet{Lotz_2008ApJ...672..177L} to classify galaxies according to their morphology. Galaxies above the solid black line are considered as merger candidates. The area above the dashed black line is usually occupied by early-type galaxies, whereas late-type galaxies are located in the region below the dashed black line. 

\begin{figure}[t]
    \centering
    \includegraphics[width=\columnwidth]{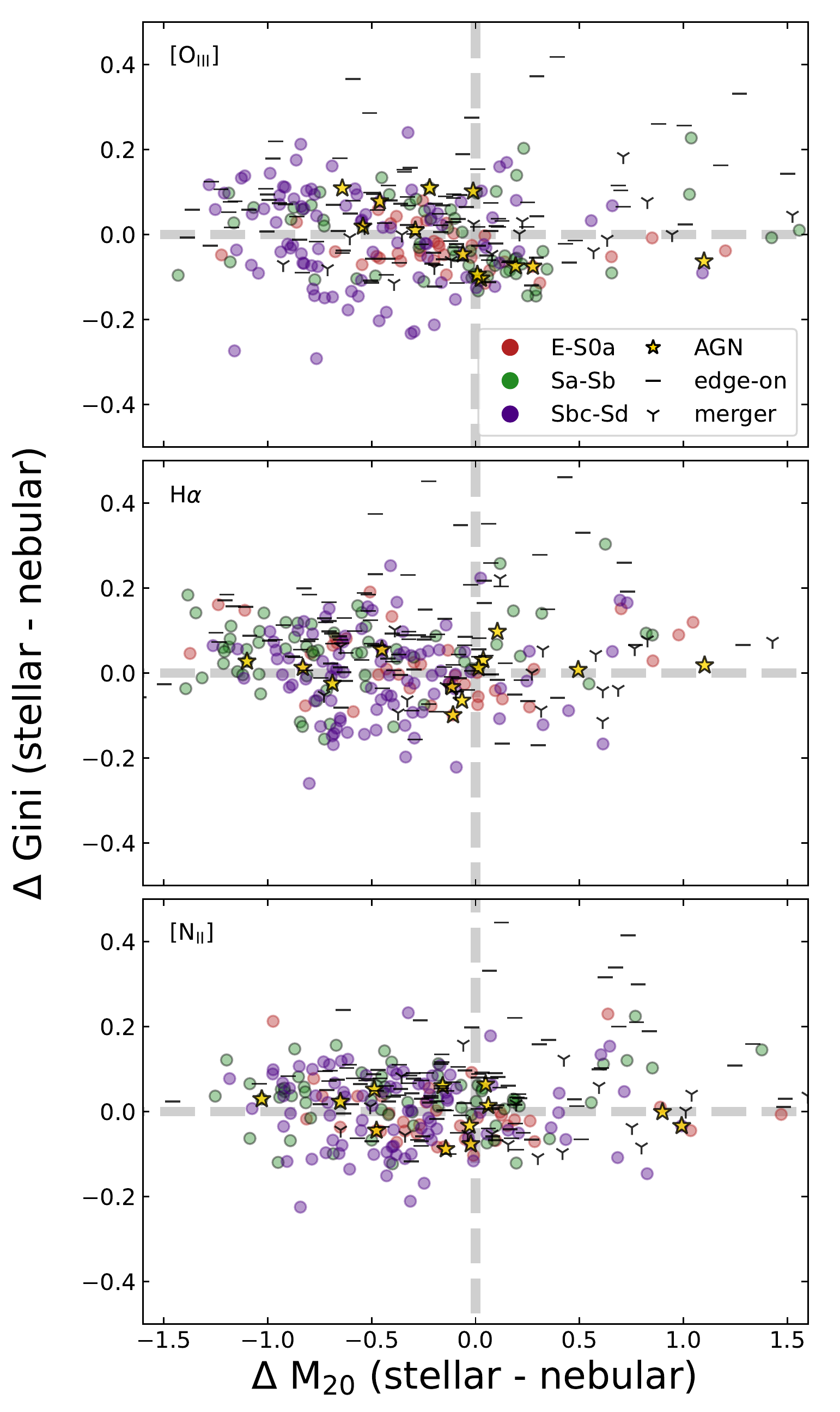}
    \caption{The change in the M$_{20}$--Gini relation from the stellar to the nebular components. Each point represents a galaxy, colour-coded according to their morphological group. AGNs are marked with a golden star. All galaxies with an inclination greater than $80\degr$ are marked with a horizontal bar, while mergers are indicated with a triangle.}
    \label{fig:delta_gini_m20}
\end{figure}

According to the morphological parameters measured from the stellar continuum we retrieve a pretty clear separation between E--S0a/Sa--Sb galaxies and Sbc--Sd. Indeed, the more compact objects like the early-types have a more concentrated light distribution (high Gini and low M$_{20}$ values) than the late-type spiral-galaxies. The 10 out of the 19 classified mergers in our sample fall to the region above the solid black line. We also notice that 24 out of the 122 edge-on galaxies in our sample are located, erroneously, to the same area as the mergers. It has been reported before that edge-on galaxies, as well as bursty galaxies, can contaminate the M$_{20}$--Gini plane \citep{Bignone_2017MNRAS.465.1106B}. Furthermore, the morphological statistics of AGN galaxies do not show any different characteristics compared to normal galaxies. Therefore, the AGN morphology seems to be determined mainly from the stellar component of their host galaxy. Of course a larger sample of AGNs of different types and with a significant variety of AGN fractions, is required for a proper statistical and morphological analysis.   

Looking now at the M$_{20}$--Gini relation for the individual spectral lines we notice that the Sbc--Sd galaxies appear to be more extended. This means that the nebular and diffuse gas emission is more extended compared to the stellar component. Many of the early-type spiral galaxies, Sa--Sb, now fall bellow the dashed black line and appear to have an equally extended nebular emission as the Sbc--Sd galaxies\footnote{This is especially true for the H$\alpha$ and [\ion{N}{II}] emission.}. Conversely, E--S0a galaxies remain roughly in the same locus, albeit with increased scatter. Regarding the AGNs, they do not follow any specific trend with the morphology measured from the nebular lines. Furthermore, all nebular lines seem to be in agreement with each other, as expected since their emission originates from the same regions within the galaxies.    

Figure~\ref{fig:delta_gini_m20} illustrates better the change in the M$_{20}$--Gini plane between the stellar continuum emission and the nebular component. In all three cases, the majority of late-type galaxies are located in the top-left quadrant. Only in the case of [\ion{O}{III}], the Sa--Sb galaxies appear to remain as concentrated as their stellar continuum, albeit with an increased scatter. On average, the M$_{20}$ and Gini index calculated from the nebular emission maps are $\sim 0.3$~dex higher and $\sim 0.03$~dex lower from the corresponding M$_{20}$ and Gini index of the stellar component. Once again, M$_{20}$ appears to be more sensitive to the nebular emission. The reason is that M$_{20}$ traces the the brightest regions outside the centre of the galaxy. The larger the value of M$_{20}$, the further away the brightest pixels are from the centre of the galaxy. In other words, the distribution of the nebular emission in late-type galaxies is more extended than their stellar disc. This effect can contaminate measurements of nonparametric indices based on broadband images. Contrarily, the nebular component of the E--S0a galaxies, when present, appears to be as compact as the stellar emission. 

\section{Discussion} \label{sec:discussion}

In this section we discuss the results of our analysis. First, we take a closer look at the wavelength dependence of the physical sizes of galaxies, averaged across various subsets of galaxy populations (subsection~\ref{subsec:galaxy_sizes}). Then, we explore how the observed inclination of galaxies affects their measured physical size and concentration index across the optical wavelength range (subsection~\ref{subsec:incl_effects}). Last but not least, we consider trends in the morphological parameters with galaxy physical properties, both for the stellar continuum and for individual spectral features (subsection~\ref{subsec:eline_morph_params_phys}).

\subsection{Galaxy size vs wavelength} \label{subsec:galaxy_sizes}

Over the years, many studies have shown the relation between galaxy size with observed wavelength. \citet{La_Barbera_2010MNRAS.408.1313L} showed a 35\% decrease in the effective radius of early-type galaxies from optical ($g$-band) to near-infrared ($K$-band) wavelengths. \citet{Kelvin_2012MNRAS.421.1007K} extended the study of \citet{La_Barbera_2010MNRAS.408.1313L} by including late-type spirals, and found a 25\% decrease in their effective radius over the same wavelength range. \citet{Vulcani_2014MNRAS.441.1340V} found that all galaxies regardless of their morphology show a decrease in size towards redder wavelengths. \citet{Mosenkov_2019A&A...622A.132M} reported a steady increase in effective radius of the dust with increasing far-infrared wavelengths, highlighting the cold-dust gradient with galactocentric distance. \citet{Baes_2020A&A...641A.119B} performed, for the first time, a consistent analysis of the morphological structure of nine late-type DustPedia galaxies \citep{Davies_2017PASP..129d4102D} as a function of wavelength, from ultraviolet (UV) to far-infrared wavelengths. They showed a decrease in galaxy size from UV to NIR, a flattening in the mid-infrared (MIR), and then an increase in the FIR, reaching a comparable size as those measured in the optical bands. In other words, they found that the distribution of interstellar dust in galaxies is as extended as the light distribution by the young stars.

We focus on the dependence of galaxy size with wavelength in the optical regime. But here we present, for the first time, a high spectral resolution, size-wavelength relation for a statistically significant sample of galaxies. The interesting part of this exercise, is that we can  eliminate the effects of line emission and isolate the size of the stellar continuum only, something that is not possible when measuring galaxy sizes from broadband images. We would like to mention again that at any given wavelength, the size $R_\mathrm{half}$ is defined as the semi-major axis of the elliptical aperture of the isophote containing half the light at that wavelength.

\begin{figure}[t]
    \centering
    \includegraphics[width=\columnwidth]{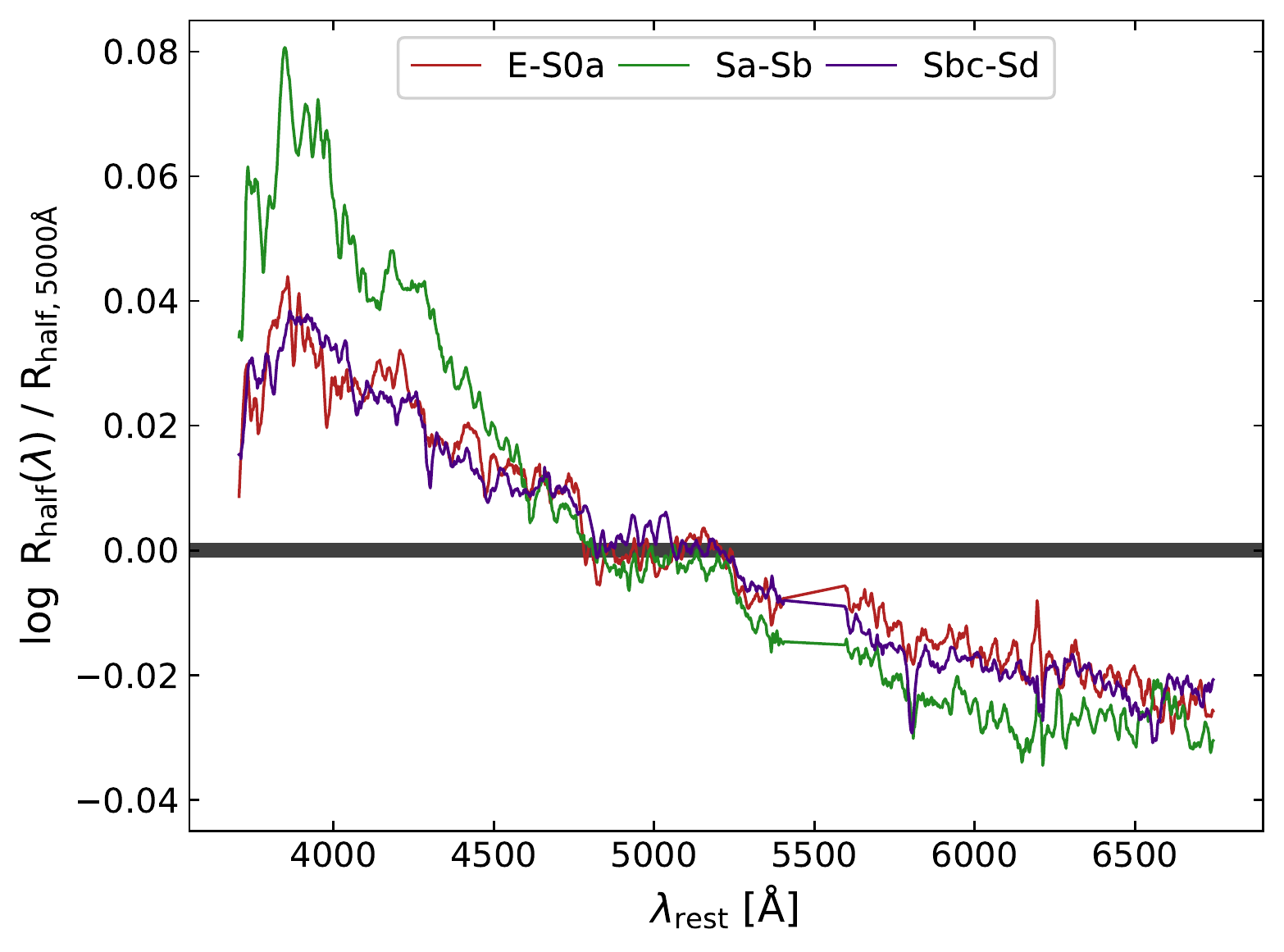}
    \caption{Normalised half-light radius $R_\mathrm{half}$ as a function of rest-frame wavelength. We bin the galaxies according to their morphological type, and estimate the median spectrum of $R_\mathrm{half}$. Then, we normalise with $R_\mathrm{half,~5000\AA}$. The spectra were binned in wavelength to smooth out the noise.}
    \label{fig:galaxy_size_vs_wavelength}
\end{figure}

We bin the galaxies into the three main morphological groups, E--S0a, Sa--Sb, and Sbc--Sd. Figure~\ref{fig:galaxy_size_vs_wavelength} depicts the logarithm of the median trends of $R_\mathrm{half}$ normalised to $R_\mathrm{half,~5000\AA}$ as a function of wavelength, and for each galaxy population. All three galaxy populations follow a similar behaviour. At short wavelengths, around 4000~$\AA$, a peak in size is reached. After this peak in size, a decrease follows with increasing wavelength, and beyond 5000~$\AA$ the size-wavelength relation becomes flatter.

What seems to be related to the morphological type, is the shape of the size-spectrum and the strength of the decrease. Below 5000~$\AA$, Sa--Sb galaxies show the most extreme relative change, since their size decreases steeply about 20\% ($\sim 0.08$~dex). The decrease in median size both for E--S0a and Sbc--Sd galaxies is smoother. The relative change in median size for the Sbc--Sd group is about 8\% ($\sim 0.035$~dex), while for the E--S0a is about 10\% ($\sim 0.04$~dex). Beyond 5000~$\AA$, the size decreases more slowly with increasing wavelength, essentially becoming flat. E--S0a have the slowest change, while Sa--Sb show again the steepest change. 

The fact that late-type spirals (Sbc--Sd) have flatter gradients than the early-type spirals (Sa--Sb), provides evidence of the lack of an old bulge in late-type spirals. In other words, it is mainly a central deficit of old stars that make Sbc--Sd galaxies to have a flatter size spectrum with respect to Sa--Sb galaxies. Early-type spirals, instead, have already formed and have largely quiescent bulges, with star formation continuing only in the disk. In addition, the similarity between E-S0a and Sbc-Sd curves indicates a more uniform surface stellar distribution while the change in Sa-Sb galaxies is due to the presence of a more extended bulge.

Another effect that may be responsible for the shape of the size-spectrum and the strength of its decrease is dust attenuation. The effect of dust is stronger in the blue wavelengths, due to the comparable dust grain size with the photon wavelength. Dust particles either absorb or scatter the blue starlight away from the line-of-sight. Therefore, galaxies might appear smaller or larger in size depending on which wavelength regime we are observing at. Lastly, the effects of dust might be stronger depending on the inclination of the system that we observe, as the column density of the dust increases with increasing inclination. 

\subsection{Inclination effects} \label{subsec:incl_effects}

\begin{figure}[t]
    \centering
    \includegraphics[width=\columnwidth]{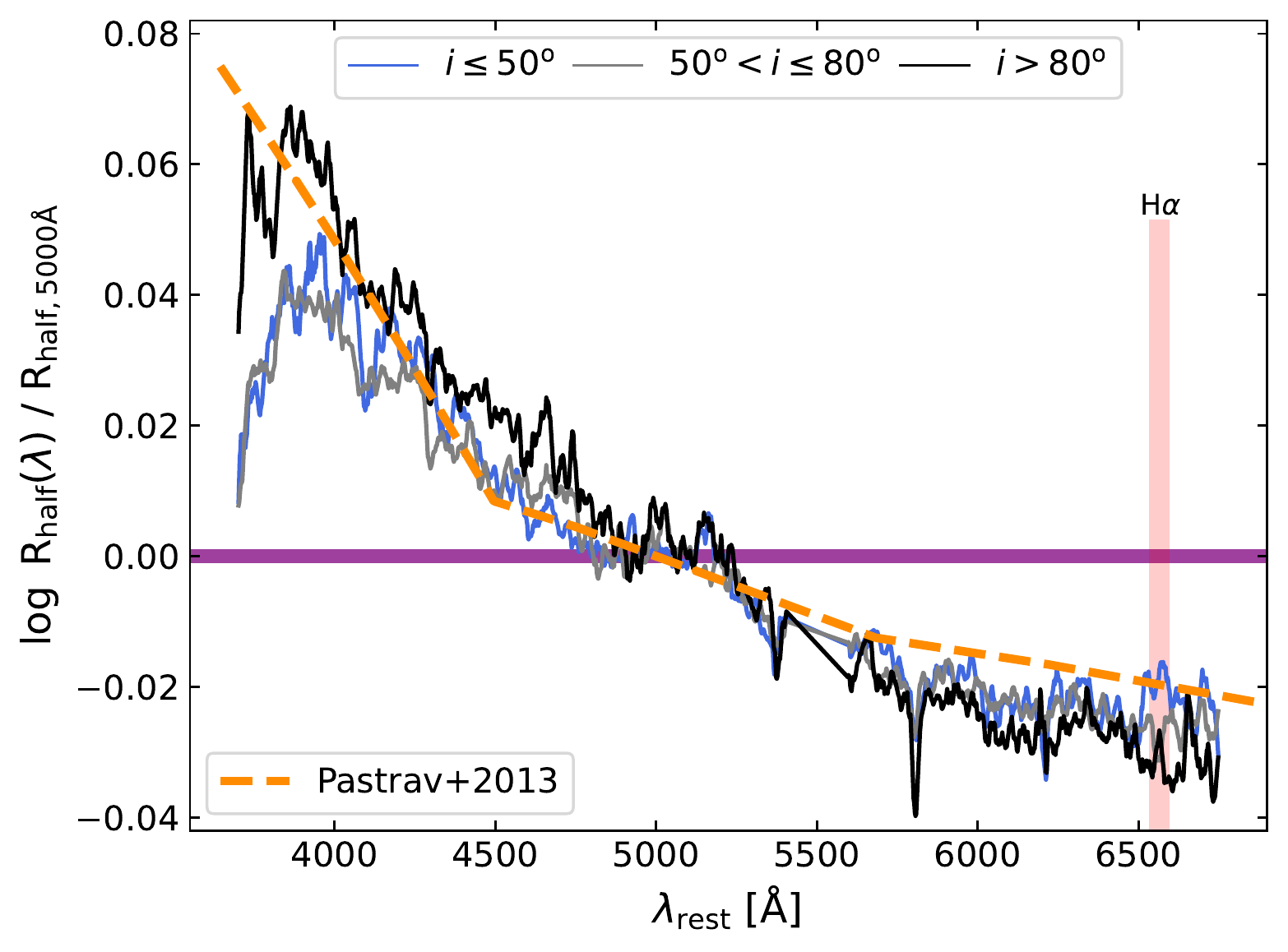}
    \caption{Normalised half-light radius $R_\mathrm{half}$ as a function of rest-frame wavelength. We bin the galaxies according to their inclination, and estimate the logarithm of the median spectrum of $R_\mathrm{half}$. Then, we normalise with $R_\mathrm{half,~5000\AA}$. We exclude the elliptical galaxies from this plot, as the effects by dust are negligible. The dashed orange line represents the dependence of size on wavelength for a disk population, and by taking into account the effects of dust. The relation was predicted by \citet{Pastrav_2013A&A...553A..80P}. The vertical red line indicates the location of the H$\alpha$ emission line.}
    \label{fig:galaxy_size_vs_wavelength_and_incl}
\end{figure}

In Fig.~\ref{fig:galaxy_size_vs_wavelength} we presented the existence of a negative gradient between the physical size of galaxies with increasing wavelength. We also noticed that the change is more drastic in some galaxies than others. The explanation is a mix of two effects: the intrinsic age and metallicity gradients of the stellar populations, and dust attenuation at short wavelengths. Yet, the fraction of stellar radiation that is reprocessed by dust, does not only depend upon the inherent properties of dust grains \citep{Draine_Li_2007ApJ...657..810D}, but also the geometry of the host galaxy \citep{Gordon_2001ApJ...551..269G, Chevallard_2013MNRAS.432.2061C}. In other words, dust attenuation should be more prominent for the more inclined disk galaxies \citep[e.g.][and references therein]{Giovanelli_1994AJ....107.2036G, Battisti_2017ApJ...851...90B, Salim_2018ApJ...859...11S, Yuan_2021ApJ...911..145Y, Doore_2021ApJ...923...26D}.

In order to better understand the variation in size, we explore another dimension in the size-wavelength relation by binning galaxies according to their observed inclination. Moreover, the dust effects are more severe for late-type galaxies than for the early-types. In the local Universe, on average, the fraction of bolometric luminosity absorbed by dust in elliptical galaxies is $\sim2\%$, for S0 is $\sim9\%$, for Sa--Sb is $\sim25\%$, while for Sbc-Sd galaxies can range from $23\%$ to $33\%$ \citep{Bianchi_2018A&A...620A.112B, Nersesian_2019A&A...624A..80N}. Hence, for this exercise we exclude the ellipticals, as the dust effects in those systems are negligible. We use three bins, separating galaxies into face-on with $i\leq50\degr$ (57 galaxies), edge-on $i>80\degr$ (111), and a bin with intermediate inclinations, $50\degr<i\leq80\degr$ (144). 

\begin{figure}[t]
    \centering
    \includegraphics[width=\columnwidth]{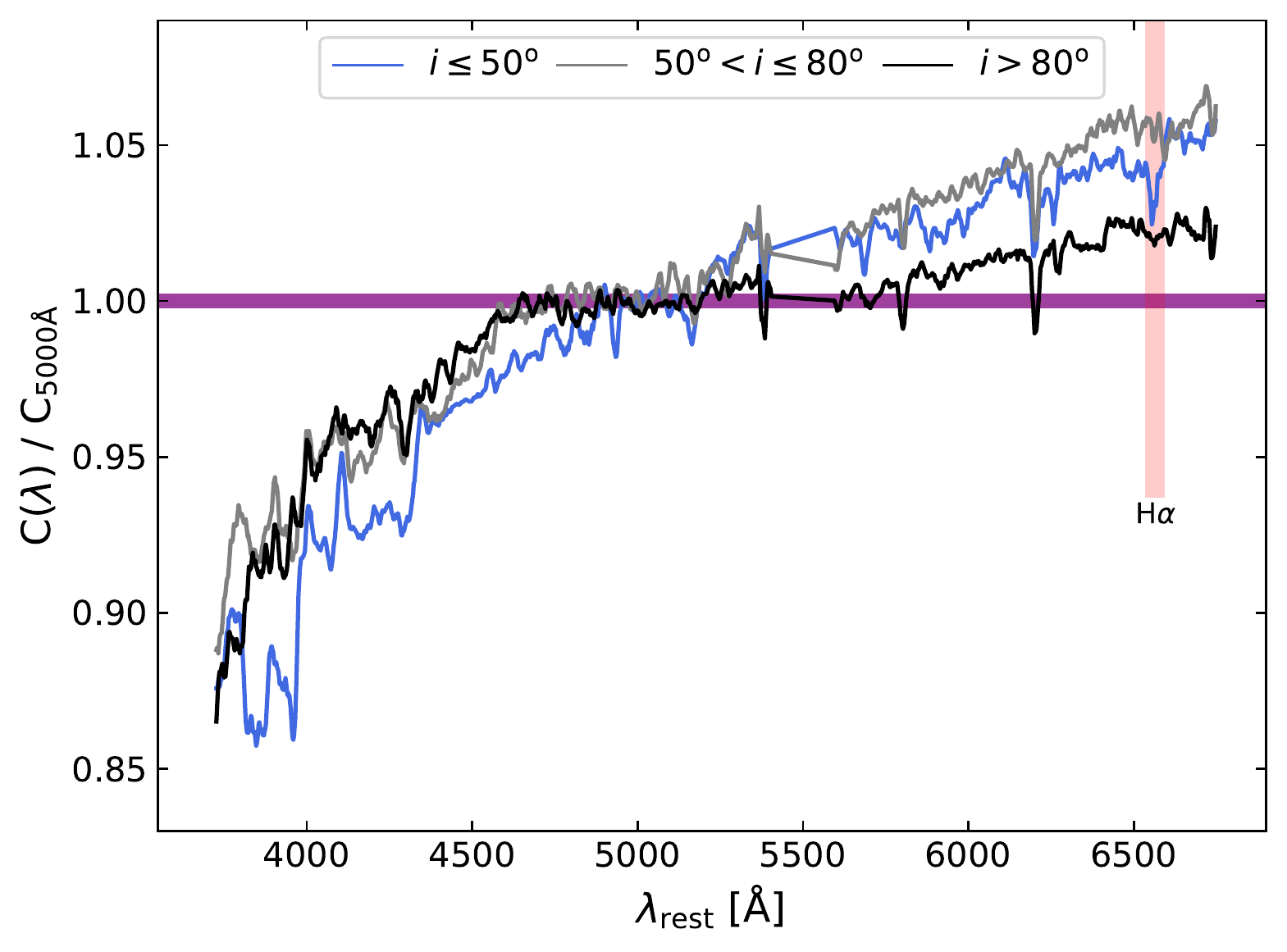}
    \caption{Normalised concentration index $C$ as a function of rest-frame wavelength. We bin the galaxies according to their inclination, and estimate the median spectrum of $C$. Then we normalise with $C_\mathrm{5000\AA}$. Similarly to Fig.~\ref{fig:galaxy_size_vs_wavelength_and_incl}, we exclude the elliptical galaxies from this plot, as the dust effects are negligible. The vertical red line indicates the location of the H$\alpha$ emission line.}
    \label{fig:concentration_vs_wavelength_and_incl}
\end{figure}

Figure~\ref{fig:galaxy_size_vs_wavelength_and_incl} shows the median trends of the $R_\mathrm{half}$, normalised to the $R_\mathrm{half,~5000\AA}$. We also plot size-wavelength relation predicted by \citet{Pastrav_2013A&A...553A..80P}. That relation was computed for a disk population and by taking into account the effect of dust attenuation. Regarding the edge-on galaxies, their physical size decreases drastically with increasing wavelength, compared to galaxies with low and intermediate inclinations where we see a smoother transition. In particular, there is a decrease of 18\% ($\sim 0.07~$dex) in size for edge-on galaxies, from 4000~$\AA$ to 5000~$\AA$. For the face-on and intermediate inclination galaxies, we see a slower change in size of the order of 11\% ($\sim 0.045~$dex), at the same wavelength range. The rapid change of the physical size of edge-on galaxies at the shortest wavelengths can be attributed to the increase in optical depth. When observing edge-on galaxies the column density of dust is increasing along the line-of-sight, either absorbing or scattering away the optical starlight. 

\begin{figure*}[t]
    \centering
    \includegraphics[width=\textwidth]{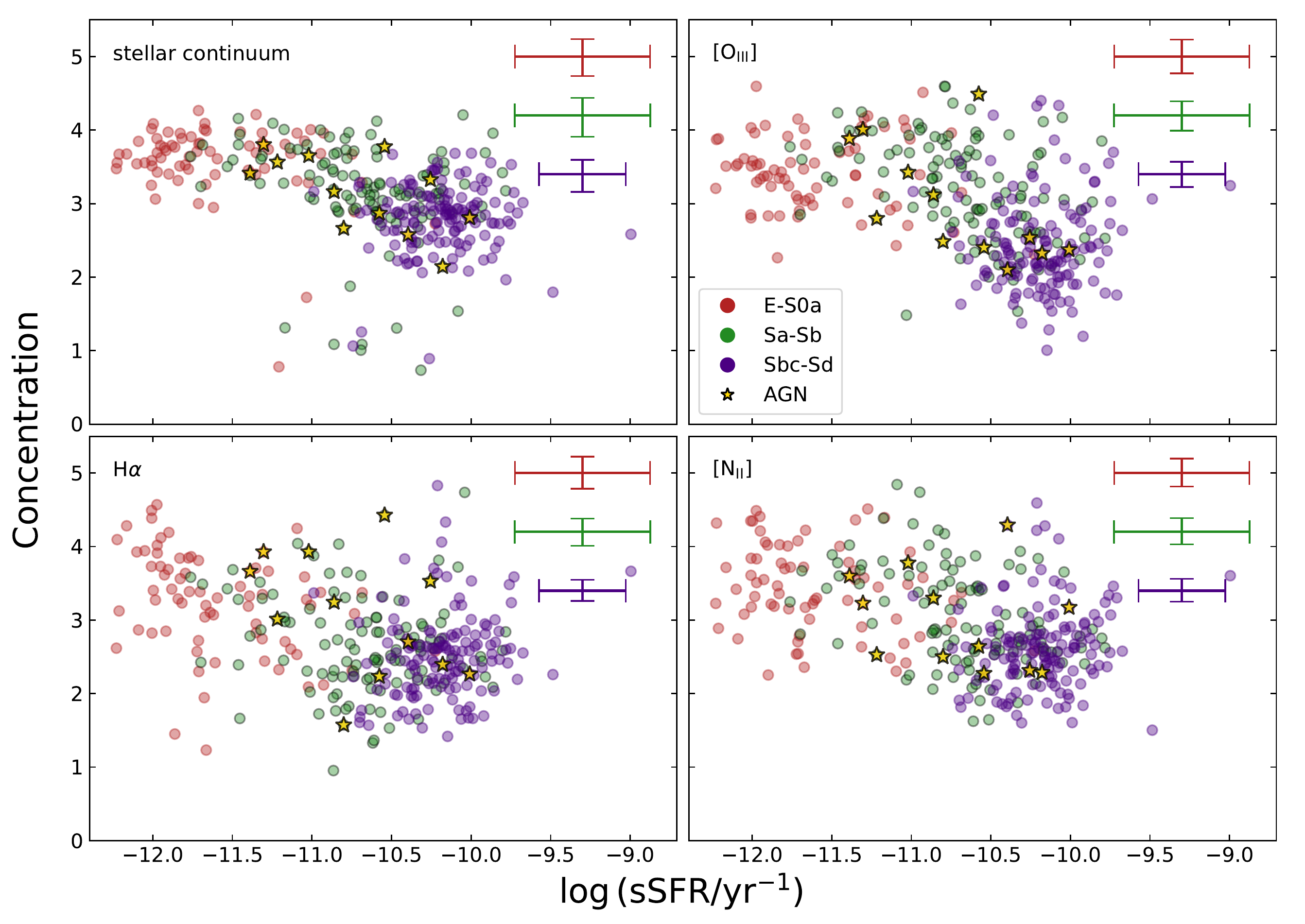}
    \caption{Concentration as a function of sSFR. Each panel shows the scatter plot of the sSFR--$C$ relation for the stellar continuum emission and three spectral lines. Each point represents a galaxy, colour-coded according to the morphological group they are belong to. AGNs are marked with a golden star. The average uncertainties of $C$ for the different galaxy groups are shown in the top-right corner of each panel. We calculate the standard deviation of sSFR for each galaxy population.}
    \label{fig:ssfr_vs_concentration_el}
\end{figure*}

\begin{figure*}[t]
    \centering
    \includegraphics[width=12cm]{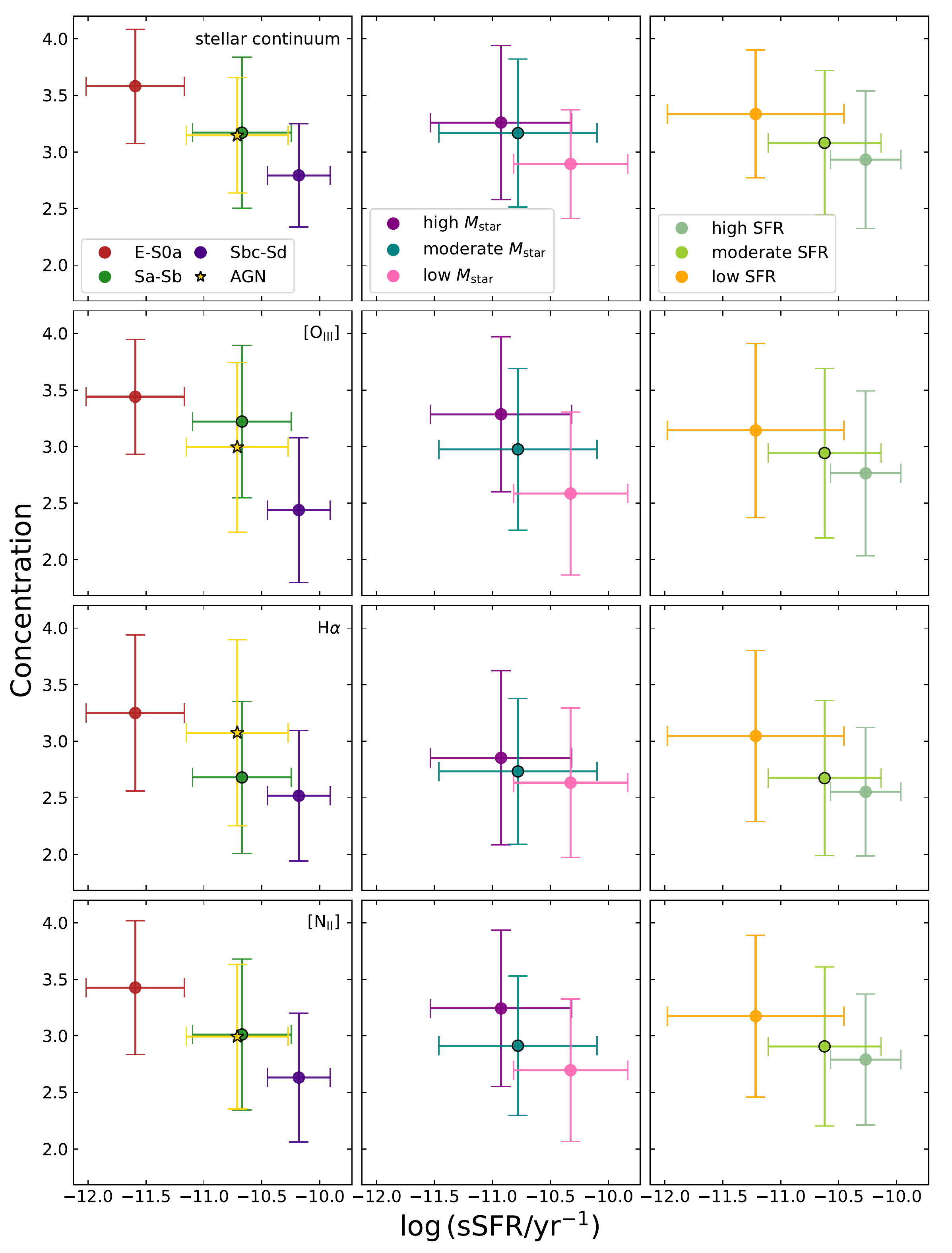}
    \caption{Average trends of the concentration index as a function of sSFR. Each row shows the scatter plot of the sSFR--C relation for the stellar continuum emission and three different spectral lines. (Left column) The data are binned according to the morphological group they are belong to, similarly to Fig.~\ref{fig:ssfr_vs_concentration_el}. (Middle column) Galaxies are binned according to their stellar mass into low $M_\mathrm{star}$ ($\log$~$M_\mathrm{star}$~$<10.4$), moderate $M_\mathrm{star}$ ($10.4\le \log~M_\mathrm{star} < 10.82$), and high $M_\mathrm{star}$ ($\log$~$M_\mathrm{star}$~$\ge10.82$). (Right column) Galaxies are binned according to their star-formation activity into low SFR ($\log$~SFR~$<-0.52$), moderate SFR ($-0.52\le \log$~SFR~$< 0.14$), and high SFR ($\log$~SFR~$\ge0.14$). All bins were defined in such a way so that they contain roughly one third of the objects in our sample. The error bars are the standard deviation of the mean values.}
    \label{fig:c_ssfr_in_bins}
\end{figure*}

Relative to the $R_\mathrm{half,~5000\AA}$, the measured half-light size around $\sim4000~\AA$ for the face-on galaxies is smaller by $\sim0.03$~dex compared to edge-on galaxies. The relative increase in size, from face-on to edge-on, is related to the tracing of the spatial extent of the stellar population, combined with a decrease of light concentration in the central regions of galaxies, where the dust density is higher. Our results are consistent with previous studies \citep[e.g.][]{Mollenhoff_2006A&A...456..941M, Leslie_2018A&A...615A...7L}, who reported an increase in the B-band sizes from face-on to an edge-on disk orientation. At wavelengths beyond 5000~$\AA$, we notice a flattening in $R_\mathrm{half}$ regardless of the observed inclination. 

In some cases, the size of emission lines such as H$\alpha$ is more extended than the stellar continuum. Primarily, the ionised emission lines originate from \ion{H}{II} regions, tracing the ongoing star formation. In Fig.~\ref{fig:galaxy_size_vs_wavelength_and_incl} we see that the size of H$\alpha$ emission in face-on galaxies is increased compared to galaxies of higher inclination, due to the fact that in face-on galaxies we see a larger fraction of the star-forming regions in the disk. 

Our results are also in a very good agreement with the predicted trend by \citet{Pastrav_2013A&A...553A..80P} for a population of disks affected solely by dust. However, the predicted relation is flatter at longer wavelengths than the observations. \citet{Pastrav_2013A&A...553A..80P} found a similar discrepancy when they compared their results with the observed trend by \citet{Kelvin_2012MNRAS.421.1007K}.  \citet{Pastrav_2013A&A...553A..80P} attributed this discrepancy to the intrinsic stellar gradients. Furthermore, \citet{Pastrav_2013A&A...553A..80P} have shown that the effects of dust on the structural properties measured in the optical wavelengths, become progressively more severe with increasing inclination. We reach the same conclusion from Fig.~\ref{fig:concentration_vs_wavelength_and_incl}, where we plot the median trends of the concentration index, as function of wavelength and in inclination bins. We normalise $C$ to the $C_\mathrm{5000\AA}$. We find that the concentration of galaxies monotonically increases with wavelength.

At wavelengths between 4000--5000~$\AA$, the light distribution of face-on galaxies is less concentrated than edge-on or highly inclined galaxies. In the case of the edge-on galaxies, the change in $C$ at the same wavelength range is more rapid, as they appear more concentrated with increasing wavelength. As the wavelength increases beyond 5000~$\AA$, the concentration index starts to diverge and to take larger values, i.e. galaxies appear more compact at redder wavelengths. The rate of change in concentration seems to depend on the observed inclination. Overall, we find that the slope becomes steeper with increasing inclination. At wavelengths greater than 5000~$\AA$, the $C$ of edge-on galaxies remains roughly constant, whereas the $C$ of face-on galaxies seems to increase.  

Lastly, once more we consistently measure that the light distribution of the H$\alpha$ emission is less concentrated than the stellar continuum, especially evident for the face-on galaxies.

Our analysis confirms that dust effects are increasingly more severe for more highly inclined galaxies. Depending on which wavelength we are observing at, the internal geometry of galaxies can have a significant impact on the measurement of their physical size and morphological properties.

\subsection{Nonparametric morphology vs physical properties} \label{subsec:eline_morph_params_phys}

Another interesting topic we want to explore is the relation between morphological indicators and various physical properties, such as SFR, stellar mass, or specific star-formation rate (sSFR). The reason is that those physical properties exhibit strong trends with morphology \citep[e.g.][]{Kennicutt_1998ARA&A..36..189K, Strateva_2001AJ....122.1861S, Schawinski_2014MNRAS.440..889S}. In this section we present the best correlation among the physical properties of the CALIFA galaxy sample and the nonparametric properties we measure from the stellar and nebular components. Using a correlation matrix and Pearson's correlation coefficient ($\rho$), we quantified the correlations between all of the products in our analysis, and find that the sSFR has the strongest correlations with the concentration index $C$. The correlation remains moderately strong with $C$ for both the stellar ($\rho = -0.47$) and nebular emission ([\ion{O}{III}]: $\rho = -0.50$; H$\alpha$: $\rho = -0.40$; [\ion{N}{II}]: $\rho = -0.47$).

The relation between the concentration index and sSFR is illustrated in Fig.~\ref{fig:ssfr_vs_concentration_el}. From this figure we start to notice differences in the galaxy morphology depending on which spectral feature we are looking at. From the sSFR--$C$ relation of the stellar continuum emission we retrieve a tight correlation with morphology. Galaxies of a certain morphological type reside in a specific location in the sSFR--$C$ plane. Of course, this trend is due to the close correlation between morphological type and sSFR \citep[e.g.][]{Bait_2017MNRAS.471.2687B, Eales_2017MNRAS.465.3125E}. E--S0a galaxies are populated by an older, more dense and compact stellar population. Their light distribution is concentrated in fewer pixels closer to their nucleus. The intermediate early-type spirals, Sa--Sb galaxies, show a decrease in concentration with increasing sSFR, as they are actively star-forming and therefore having more extended emission than the non-star-forming E--S0a galaxies. Lastly, the late-type spirals, Sbc--Sd, are those with large star-forming disks having the lowest concentration and highest sSFR (high SFR and low stellar mass).

When looking at the same relation measured from the three emission lines we notice a considerable increase in the scatter. However, galaxies still remain well separated according to their morphological group. An interesting point to notice here is that when the $C$ is measured from the stellar continuum it is only possible to separate the E-S0a from the late-type spirals, since both Sa-Sb and Sbc-Sd types have very similar values. A better separation among the morphological types can be achieved when measuring the $C$ from the [\ion{O}{III}] emission line.

Moreover, the [\ion{O}{III}], H$\alpha$, and [\ion{N}{II}] show a negative correlation between sSFR and concentration, similar to the continuum emission. Yet again, the nebular emission seems to be less concentrated than the stellar continuum, indicating the extended distribution of the star-forming regions in the disks of galaxies. The concentration of the emission lines in E--S0a is remains roughly constant compared to the stellar continuum. This gives support to the idea that in early-types any of their remaining ionised gas is mostly concentrated to their central regions.  

In Fig.~\ref{fig:c_ssfr_in_bins} we show the concentration index measured from the stellar continuum and the three most prominent spectral features in the CALIFA data cubes, as a function of sSFR. We separate the data into morphological (left column), mass (middle column), and SFR (right column) bins. We are using the same morphological bins that have been used throughout the paper (excluding the 37 \texttt{flagged} galaxies), as well as a bin that includes all AGNs (12 galaxies) in the sample. For the stellar mass we are using three bins, one of low $M_\mathrm{star}$ ($\log$~$M_\mathrm{star}$~$<10.4$), moderate $M_\mathrm{star}$ ($10.4\le \log~M_\mathrm{star} < 10.82$), and high $M_\mathrm{star}$ ($\log$~$M_\mathrm{star}$~$\ge10.82$). Similarly we have three SFR bins, one of low SFR ($\log$~SFR~$<-0.52$), moderate SFR ($-0.52\le \log$~SFR~$< 0.14$), and high SFR ($\log$~SFR~$\ge 0.14$). All bins were defined in such a way so that they contain roughly one third of the objects in our sample.

The left column of Fig.~\ref{fig:c_ssfr_in_bins} presents the average trends of the scatter plots in Fig.~\ref{fig:ssfr_vs_concentration_el}. Immediately we notice that $C$ has a decreasing trend with sSFR and that galaxies are clearly separated into early-, intermediate-, and late-types. The same trend can be seen whether we measure the concentration index from the stellar continuum or from the nebular emission lines. On average, early-type galaxies have higher concentration indicating a more compact light distribution. Intermediate galaxies decrease in concentration but overall take similar values as the early types, nevertheless they are less concentrated. The late-type spirals show low concentration i.e. they have a more extended light distribution, as expected, since they are disk dominated systems. 

We find that the mean value of the concentration of early-types remains roughly constant in all panels. The relative change in $C$ between the stellar continuum and the nebular lines is less than $11\%$. In late-type galaxies the nebular emission appears to be more extended than the stellar continuum. In particular, for Sa--Sb the mean value of $C$ of the stellar continuum is consistent with the [\ion{O}{III}] emission, while the H$\alpha$ and [\ion{N}{II}] emission is less concentrated than the stellar continuum by 15.6\% and 6.3\%, respectively. For the Sbc--Sd types, we find a difference of 14.3\%, 10.7\%, and 7\% for the [\ion{O}{III}], H$\alpha$ and [\ion{N}{II}] emission lines, respectively.

In terms of sSFR, galaxies that host an AGN occupy the same locus as the Sa--Sb types, suggesting that AGNs are in a transitioning phase from actively star-forming to quiescence. This is in agreement with previous studies reporting that AGN galaxies have analogous SFRs with non-AGN galaxies of the same evolutionary stage \citep{Suh_2017ApJ...841..102S, do_Nascimento_2019MNRAS.486.5075D}. In AGNs the relative change of the mean $C$ as measured from the stellar continuum and the nebular emission lines is rather small ($< 3.3\%$). Interestingly enough, the $C$ measured from H$\alpha$ shows a difference from the corresponding value of Sa--Sb galaxies (although the standard deviation is high). On average, it appears that the H$\alpha$ emission of AGN galaxies is $13\%$ more compact compared to the Sa--Sb galaxies. The most plausible explanation of this result is the extra nuclear emission due to the AGN. Of course, we would like to stress here that the trends we find for AGN have large uncertainties due to the small sample size. Further, investigation with a larger sample is needed to draw robust conclusions. 

The middle column of Fig.~\ref{fig:c_ssfr_in_bins} shows the $C$ as a function of sSFR, binned by stellar mass. We immediately notice that the more massive galaxies are also the more concentrated. Then, the mean $C$ values decrease with increasing sSFR and decreasing stellar mass. By inspecting the right column of Fig.~\ref{fig:c_ssfr_in_bins} we see that the mean $C$ values decrease as the sSFR increases and the SFR of the galaxies becomes larger. From this we conclude that the more star-forming a galaxy is, the less concentrated it becomes, because star formation is distributed on disks that are generally larger than the pre-existing stellar distribution. The same conclusions can be reached when looking at the mean trends for the nebular emission lines.

In general, we conclude that the apparent morphology of galaxies originates from the starlight distribution, and the morphology of the nebular component differs only moderately from the morphology of the stellar body. Primarily, the emission of the different nebular lines in late-type galaxies probe the light from [\ion{H}{II}] regions, tracing the spatial extent of the ongoing star formation in their disk.

\subsection{Future applications} \label{subsec:outro}

The results of this work provide a baseline for more extensive studies both on observational and simulated data. A straightforward application of our methodology could be performed on other integral field spectroscopic surveys such as PHANGS-MUSE \citep{Emsellem_2022A&A...659A.191E} that contains higher spatial resolution, or MaNGA \citep{Bundy_2015ApJ...798....7B} and SAMI \citep{Bryant_2015MNRAS.447.2857B}, that contain larger galaxy samples. Performing a similar analysis, but for a larger statistical sample, will allow to further establish the reported morphological trends in our study.  

Furthermore, our results can be leveraged as benchmark for models of galaxy formation and evolution. The degree to which cosmological hydrodynamical galaxy simulations agree with our results is a major test for them. Specifically, the morphology of galaxies is something that is not considered in the calibration of the models, unlike, for example, the galaxy stellar mass function \citep[e.g.][]{Vogelsberger_2014Natur.509..177V, Schaye_2015MNRAS.446..521S, Pillepich_2018MNRAS.473.4077P, Dave_2019MNRAS.486.2827D}. Recently, \citet{Kapoor_2021MNRAS.506.5703K} and \citet{Camps_2022MNRAS.512.2728C} provided comparisons of the broadband UV-submm morphological indicators of Auriga \citep{Grand_2017MNRAS.467..179G} and ARTEMIS \citep{Font_2020MNRAS.498.1765F, Font_2021MNRAS.505..783F} simulated galaxies to those from DustPedia observations. In both studies, the authors found that the observed morphological trends as a function of wavelength are in a reasonable agreement with those measured from simulated galaxies. Our work enables for a more detailed comparison of both the stellar continuum and the nebular emission.

\section{Summary and Conclusions} \label{sec:conclusions}

In this paper, we presented a nonparametric morphology analysis of the stellar continuum and nebular emission lines with a subset of CALIFA galaxies. We explored the dependence of the various morphological parameters on wavelength and morphological type, and quantified the difference in morphology between the stellar and nebular components. The nonparametric morphological indicators were derived with the \texttt{StatMorph} package, applied on the CALIFA IFS data cubes, as well as to the most prominent nebular emission maps, namely [\ion{O}{iii}]$\lambda$5007, H$\alpha$, and [\ion{N}{ii}]$\lambda$6583.

We showed a strong gradient, from blue to red optical wavelengths, for the physical size of galaxies, M$_{20}$ index, and concentration. In particular, we find that M$_{20}$ is much more sensitive to the morphology of emission lines than the other nonparametric morphological indices. We find that the light distribution of the nebular emission follows the same trend as the stellar continuum, but is less concentrated. From this result we conclude that the apparent morphology of galaxies originates from the stellar light distribution, and that the morphology of the ISM show noticeable differences from the morphology of the stellar component.

A comparison between the nonparametric indicators and the galaxy physical properties, revealed a very strong correlation of the concentration index with the specific star-formation rate and morphological type. We find that early-type galaxies are more concentrated than late-types. By binning galaxies according to their stellar mass and ongoing SFR, we find that the more massive and low star-forming galaxies are more concentrated, and vice versa.

We also report evidence that the morphology of the nebular emission is more asymmetric and more clumpy than the stellar emission. We note here that the measured values for $A$ and $S$ come with large uncertainties due to the low S/N and spatial resolution of the CALIFA IFS data cubes. 

Finally, we explored how galaxy inclination affects our results. We find that high inclined galaxies show a more rapid change in physical size and concentration with increasing optical wavelength, due to the increase in optical depth. Our analysis highlights galaxy inclination as an important aspect of the measurement of the nonparametric morphological indicators, that needs to be considered when interpreting the results. 

The median trends of the nonparametric indices as a function of wavelength, for each morphological group, that are shown in Fig.~\ref{fig:medians}, and the Gini, M$_{20}$, and concentration indices of the stellar continuum and nebular emission lines that were used in the analysis of this work are available in electronic form at the Centre de Données Astronomiques (CDS) de Strasbourg. The morphological parameters as a function of wavelength of each galaxy, presented in Fig.~\ref{fig:morph_spectra}, are available from the corresponding author AN on request.

\begin{acknowledgements}
We would like to thank the referee, Jaime Perea, for the helpful comments and suggestions. AN gratefully acknowledges the support of the Research Foundation - Flanders (FWO Vlaanderen). FDE acknowledges funding through the ERC Advanced grant 695671 ``QUENCH'' and support by the Science and Technology Facilities Council (STFC). This study uses data provided by the Calar Alto Legacy Integral Field Area (CALIFA) survey (http://califa.caha.es/). Based on observations collected at the Centro Astron\'{o}mico Hispano Alem\'{a}n (CAHA) at Calar Alto, operated jointly by the Max-Planck-Institut f\H{u}r Astronomie and the Instituto de Astrof\'{i}sica de Andaluc\'{i}a (CSIC). This research made use of Astropy,\footnote{\url{http://www.astropy.org}} a community-developed core Python package for Astronomy \citep{Astropy_2013A&A...558A..33A, Astropy_2018AJ....156..123A}. 
\end{acknowledgements}




\bibliographystyle{aa}
\bibliography{References} % if your bibtex file is called example.bib

\end{document}